%% file: main.tex
\documentclass[twocolumn]{aastex631}
\usepackage{verbatim}
\usepackage{hyperref}
\hypersetup{
    colorlinks=true,
    linkcolor=blue,
    filecolor=magenta,      
    urlcolor=cyan,
    citecolor=blue,
}
\usepackage{amsmath}
\usepackage{comment}
\input{macros.tex}
\newcommand\update[1]{{#1}}
\graphicspath{{figures/}}

\shorttitle{\tess\ Discovery of Twin Planets Around \target\ }
\shortauthors{Tey et al.}

\begin{document}

\title{\tess\ Discovery of Twin Planets near 2:1 Resonance around Early M-Dwarf TOI 4342 }

\input{authors.tex}

\begin{abstract}
With data from the Transiting Exoplanet Survey Satellite (\tess), we showcase improvements to the MIT Quick-Look Pipeline (QLP) through the discovery and validation of a multi-planet system around M-dwarf \target\ ($T_{\text{mag}}=\starApproximateTMag$, \mstar\ = \starApproximateMass\ \msun, \rstar\ = \starApproximateRadius\ \rsun, \teff\ = \starApproximateTeff\ K, d=\starApproximateDistance\ pc). With updates to QLP, including a new multi-planet search, as well as faster cadence data from \tess\ ' First Extended Mission, we discovered two sub-Neptunes ($R_b=\bRadius$ \rearth\ and $R_c=\cRadius$ \rearth; $P_b=\bApproximatePeriod$ days and $P_c=\cApproximatePeriod$ days) and validated them with ground-based photometry, spectra, and speckle imaging. Both planets notably have high transmission spectroscopy metrics (TSMs) of \bApproximateTSM\ and \cApproximateTSM, making \target\ one of the best systems for comparative atmospheric studies. This system demonstrates how improvements to QLP, along with faster cadence Full-Frame Images (FFIs), can lead to the discovery of new multi-planet systems. 

% Filled out version for submission
% With data from the Transiting Exoplanet Survey Satellite (TESS), we showcase improvements to the MIT Quick-Look Pipeline (QLP) through the discovery and validation of a multi-planet system around M-dwarf TOI 4342 (T<sub>mag</sub>=11.032, M<sub>*</sub> = 0.63 M<sub>sun</sub>, R<sub>*</sub> = 0.60 R<sub>sun</sub>, T<sub>eff</sub> = 3900 K, d = 61.54 pc). With updates to QLP, including a new multi-planet search, as well as faster cadence data from TESS' First Extended Mission, we discovered two sub-Neptunes (R<sub>b</sub> = 2.266<sub>-0.038</sub><sup>+0.038</sup> R<sub>earth</sub> and R<sub>c</sub> = 2.415<sub>-0.040</sub><sup>+0.043</sup> R<sub>earth</sub>; P<sub>b</sub> = 5.538 days and P<sub>c</sub> = 10.689 days) and validated them with ground-based photometry, spectra, and speckle imaging. Both planets notably have high transmission spectroscopy metrics (TSMs) of 36 and 32, making TOI 4342 one of the best systems for comparative atmospheric studies. This system demonstrates how improvements to QLP, along with faster cadence Full-Frame Images (FFIs), can lead to the discovery of new multi-planet systems.
\end{abstract}


\section{Introduction}
\label{sec:intro}
The \textit{Transiting Exoplanet Survey Satellite} \citep[\tess,][]{tess} launched on April 18, 2018 with a goal of discovering transiting exoplanets around bright stars across the entire sky. During every sector ($\sim 27$ days), \tess\ observes a $24^{\circ} \times\ 96^{\circ}$ swath of the sky during two elongated orbits around the Earth before shifting to the next sector. During its Primary Mission (2018 Jul 25 -- 2020 Jul 04), \tess\ collected photometry at a 2 minute cadence for $\sim$ 20,000 targets each sector pre-selected from the Candidate Target List \citep[CTL,][]{stassun2018}, while Full-Frame Images (FFIs) were collected at a 30 minute cadence. In these 26 sectors, \tess\ covered 70\% of the sky and found 2241 transiting planet candidates \citep{toi}.

The MIT Quick Look Pipeline \citep[QLP,][]{qlp} has been an important contributor to detecting these planet candidates, expanding the search from the $\sim 200,000$ CTL stars to millions of stars brighter than $T_{\text{mag}}=10.5$ in the FFIs. Every sector, QLP searches for planet candidates around stars with $T_{\text{mag}}<10.5$, making full use of all past sectors of data. Notably, $\sim 1000$ candidates from the Primary Mission were found around stars \textit{not} on the CTL -- a majority of which were detected by QLP.

With its 1st Extended Mission (2020 Jul 04 -- 2022 Sep 01), \tess, and QLP in particular, was well-positioned to yield even more candidates, especially as the FFI recording cadence was reduced from 30 minutes to 10 minutes. On 2022 Sep 01, \tess\ will have started its Second Extended Mission, and the FFI recording cadence will be reduced further to 200 seconds. These reductions allow the flux-time series from the FFIs to better resolve transit ingresses and egresses and therefore improve planet detection efficiency. QLP, however, still has room for improvement. We are introducing here three major changes to the pipeline:

\begin{itemize}
    \item Systematic Multi-planet Search: In the Primary Mission, QLP only sent the most promising transit candidate from each light curve through future stages of vetting. Additional candidates could be reported if the TOI vetters noticed them during visual inspection of the light curves, but no explicit search was otherwise done for these additional planets. Based on our knowledge of close-in exoplanet systems from the Kepler mission, a majority of the small planets reside in multiple planetary systems \citep{winn2015}. Adding an automatic search for multiple transit signals will potentially introduce new interesting systems for follow-up studies and allow future statistical studies of system architecture. 
    \item Improved Difference Images: Difference images \citep{bryson2013} are an effective method of determining the source location of a transit signal in the sky. For the Primary Mission, QLP used a simplistic algorithm for creating difference images by directly subtracting the median stacked frames in the in-transit and out-of-transit time windows. By more intelligently selecting frames to avoid systematics, transit ingress / egress, and additional planets in the system, we improve the robustness of the difference images. This reduces QLP's false positive rate and allows us to reduce human-review times or alternatively expand our search to more stars. 
    \item Quaternion Detrending: Previously, QLP detrended light curves by fitting and removing basis splines to correct for long-timescale systematics \citep{vanderburg2014}. However, there are residual systematics on shorter timescales due to space craft jitter motions that can have significant effects on light curves, especially for bright stars \citep{toi-396, toi-1130}. We can measure and correct for this jitter with \tess\ quaternion data -- 3-vector time series data describing spacecraft attitude every two seconds.\footnote{Quaternion data is available online at \url{https://archive.stsci.edu/missions/tess/engineering/}} For transit signals with relatively small depth and short duration, correcting these systematics is important to improving QLP's detection sensitivity. 
\end{itemize}

Altogether, these improvements to QLP -- along with extended light curves at faster cadence -- will increase the scientific output of \tess\ as a whole. One population that will especially benefit from this are multi-planet systems around M-dwarfs. M-dwarfs are known to frequently host multi-planetary systems \citep{ballard2019, dressing2015}. The duration of planetary transits around M-dwarfs are often relatively short, making these transits easily diluted in the 30 minute cadence \tess\ FFI light curves from the Primary Mission. M-dwarfs are also the most abundant type of star in our galaxy. So, even though many have been selected for 2 minute cadence target pixel stamp observations, plenty are only monitored by the FFIs. The regular multi-sector processing approach of the QLP also enables us to take the most advantage of all the available data. 
Finally, M-dwarfs are also ideal host stars for atmospheric characterization of their transiting planets. Determining atmospheric composition, especially for multiple planets in a single system, allows us to compare formation and evolutionary histories.

As of July 2022, \tess\ has detected 373 candidate transiting planetary systems around nearby M-dwarfs \footnote{\update{The \tess\ candidate list was downloaded from the NASA Exoplanet Archive on July 7, 2022.}}. 36 of these systems host multiple planets, including TOI 270 \citep{toi-270:gunther, toi-270:vaneylen2021}, TOI 175 \citep{toi-175:kostov, toi-175:cloutier}, and TOI 700 \citep{toi-700:gilbert, toi-700:rodriguez}. Building off the original Kepler mission \update{\citep{https://doi.org/10.26133/nea5}}, this doubles the number of discovered transiting M-dwarf multi-planet systems.

The discovery of \target, an M-dwarf system with two transiting sub-Neptunes, showcases the power of the Extended Mission FFIs with the newly added improvements to QLP for detecting multi-planet M-dwarf systems. With the longer baseline, we have more opportunities to catch transits and, together with the faster cadence, we have better statistics to detect the shorter-duration transits that come with M-dwarf systems. Adding short-timescale systematic corrections adds to this sensitivity, and the multi-planet search is what allows us to discover this multi-planet system where the previous version of QLP could not make this discovery.

\target\ (TIC 354944123; $T_{mag}=\starApproximateTMag$; $d=\starApproximateDistance$ pc) is an M0V-type star with two transiting sub-Neptunes \bPlanet\ (\bApproximateRadius\ \rearth) and \cPlanet\ (\cApproximateRadius\ \rearth). The planets are near a 2:1 mean-motion resonance with periods of \bApproximatePeriod\ and \cApproximatePeriod\ days. Both planets are good targets for atmospheric characterization \& comparison studies with transmission spectroscopy metrics $\geq 30$ \citep[TSM,][]{kempton2018}.

Both planets were detected with \tess\ and followed up with ground-based photometry, reconnaissance spectroscopy, and high resolution imaging. In Section \ref{sec:data} we describe these observations. In Section \ref{sec:analysis} we perform fits and validate both signals as planetary transits around an M0V host star. In Section \ref{sec:discussion} we re-emphasize the improvements to QLP and describe the \target\ system in the context of small planets, multi-planet systems, and planets orbiting M-dwarfs.

\section{Observations and Data analysis}
\label{sec:data}
\subsection{TESS}
\label{sec:tess}

\target\ was first observed by \tess\ in Sector 13 (Primary Mission, 2019 Jun 19 – Jul 18) then observed again in Sector 27 (1st Extended Mission, 2020 Jul 4 – 30, 2020) as a 2-min target due to its brightness and small radius \citep{stassun2019}, as well as in FFIs. 

The 2-min data were reduced by the Science Processing Operations Center (SPOC) pipeline at NASA \citep{spoc}. The planets’ transit signals were detected during a SPOC multisector search of Sectors 13 and 27 on 26 May 2021 with an adaptive, noise-compensating matched filter \citep{jenkins2002, jenkins2010, jenkins2020}. In the multisector search, the 5-day signal was detected with SNR 10.8 and multiple event statistic (MES) of 8.5, and the 10-day signal was detected with SNR 9.0 and MES 8.2. Both transit signatures passed all the diagnostic tests reported in the SPOC data validation reports \citep{twicken2018, li2019}, including the difference image centroid tests, which located the source of the transits to within 8.7 +/- 5.5 arcsec and 3.4 +/- 3.6 arc sec of the target star image for \target\ b and c, respectively. \update{Both candidates were classified as high quality candidates by the TESS-ExoClass classifier \footnote{\url{https://github.com/christopherburke/TESS-ExoClass}}. TESS-ExoClass applies a series of tests based on the Kepler Robovetter \citep{2016ApJS..224...12C, 2018ApJS..235...38T} to pass the best candidates on to the manual TOI vetting process \citep{toi}}. These exoplanet signatures were alerted as \target.01 and .02 on 28 July 2021 \citep{toi}.

\begin{figure*}
    \centering
    \includegraphics[width=\textwidth]{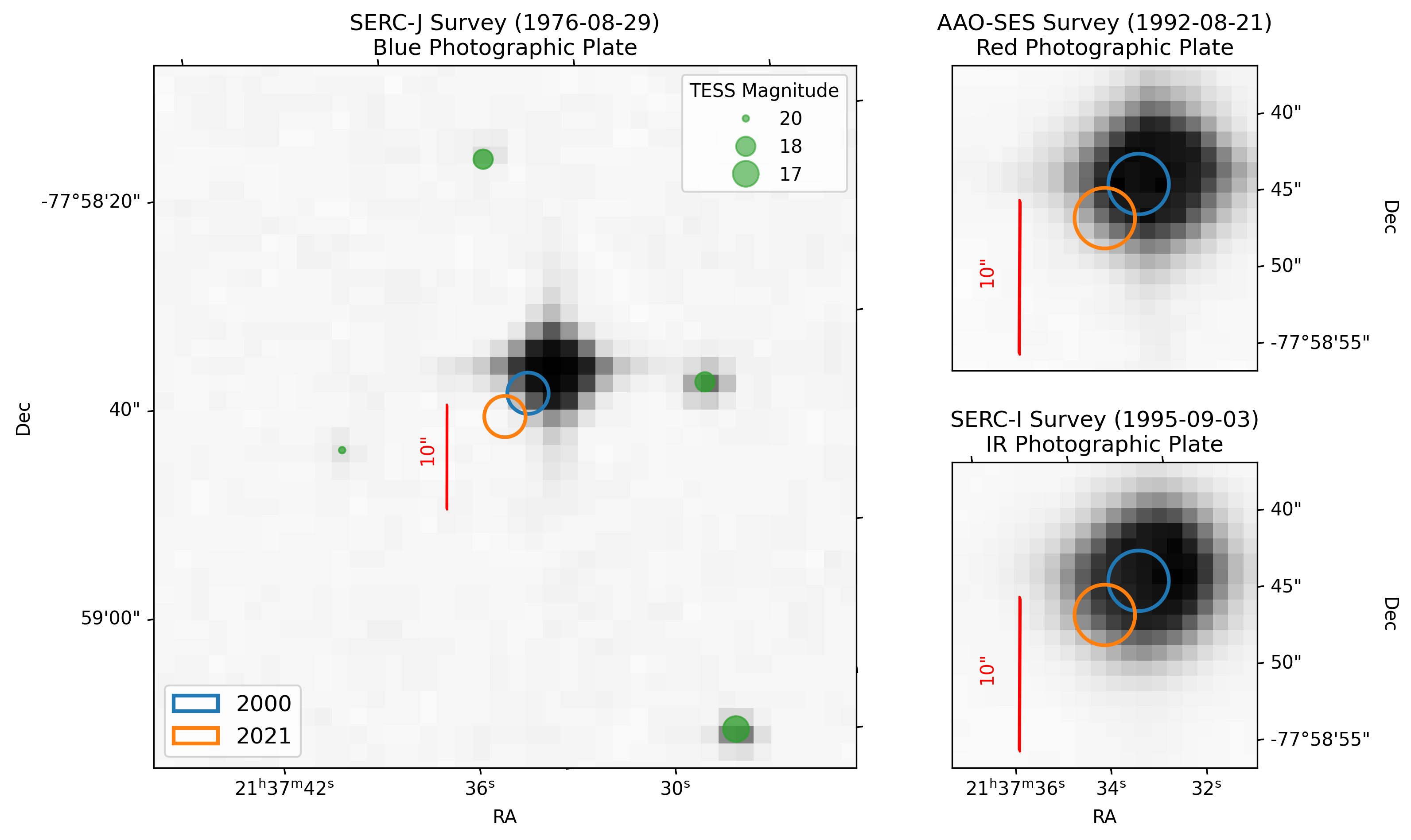}
    \caption{Images of the field surrounding \target\ in 1976 (left), 1992 (top right), and 1995 (bottom right). In all images, the orange circle shows the 2021.0 location of \target\ and the blue circle shows the location on J2000. Both circles have radii of 2\arcsec. Nearby stars in the TIC are shown in green with marker size corresponding to brightness. In red is a 10\arcsec\ scale ($\sim$ half a \tess\ pixel).}
    \label{fig:field}
\end{figure*}

The FFIs meanwhile were reduced by QLP, but \target’s \tess\ band brightness is below the threshold to be released as a QLP TOI (10.5) \citep{toi}. While developing the QLP improvements listed in Section \ref{sec:intro}, we conducted an independent search for multi-planet systems in QLP light curves and found two planet candidates in \target. For both sectors, our analysis started with raw QLP light curves, which uses calibrated FFIs from the MIT TESS image calibration software \citep[TICA,][]{tica} \footnote{TICA FFIs are available as High Level Science Products at the Mikulski Archive for Space Telescopes (MAST): \url{https://archive.stsci.edu/hlsp/tica}}. Figure \ref{fig:field} shows \target\ and the surrounding field.

This multi-planet search consisted of iterative applications of the Box Least Squares algorithm \citep[BLS,][]{bls} as implemented in \texttt{VARTOOLS} \citep{vartools}, masking out transits for previously found signals when searching for new planets. We then also performed an additional step of light curve detrending to remove short-timescale systematics. This was done by decorrelating spacecraft motion from the light curve using leading statistical moments (mean, standard deviation, skewness) of and covariances between the quaternion time series components Q1, Q2, and Q3 calculated within each exposure \citep{toi-396}. Together with basis splines to remove long-timescale trends \citep{vanderburg2014}, we perform iterative fits to the light curve, removing 3-$\sigma$ outliers until the fit converged. The resulting combined trend was subtracted out to produce a final corrected QLP light curve.

With these adjustments, we found two signals. The first signal has a period of \bQLPPeriod\ days with signal-to-noise ratio (SNR) of \bQLPSNR\ (5.14 per transit), and the second has a period of \cQLPPeriod\ days with SNR \cQLPSNR\ (6.74 per transit). To highlight the impact of the first Extended Mission and QLP improvements, we specifically searched for planets in Sector 13 (Primary Mission) and Sector 27 (first Extended Mission) separately. Corrected for the number of transits seen in each sector, we saw average SNR per transit of 3.70 and 4.83 in Sector 13 for each signal respectively. In Sector 27, this improved to 5.52 and 5.99. We also compared the SNRs with the original QLP detrending method (only correcting for long-timescale systematics) and found worse performance without the quaternion correction (SNRs per transit of 5.07 and 5.32).
% qsp: 5.14, 6.73
% ksp: 5.07, 5.32

\begin{figure*}
    \centering
    \includegraphics[width=\textwidth]{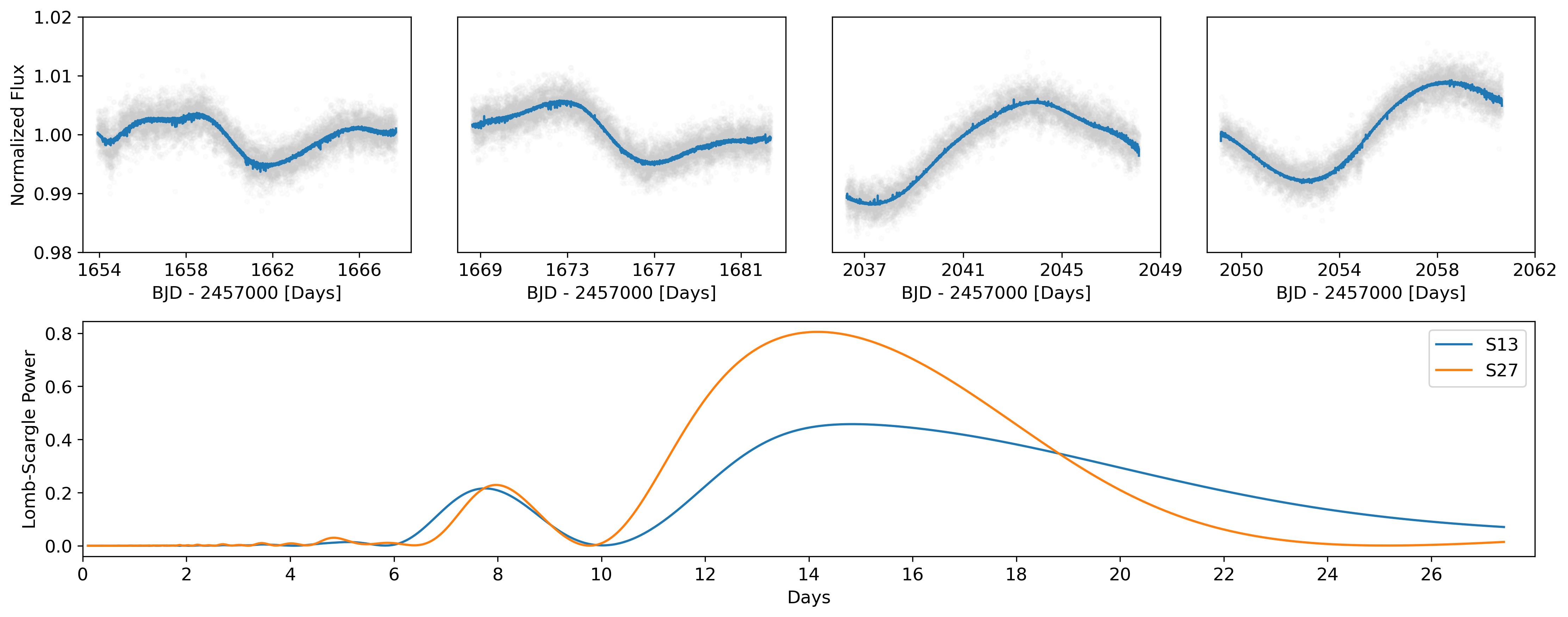}
    \caption{Top: SPOC SAP light curve with our correction trend (described in Section \ref{sec:tess}) in blue. This trend, formed from quaternion data and basis splines to remove systematics and long timescale variability, is divided out of the raw light curve to create our final flattened light curve. Bottom: Lomb-Scargle periodograms of the SAP SPOC light curve for each sector of observation. We see peaks around 14 days marking stellar variability, but we cannot pin down a specific period due to the large gap between observations.}
    \label{fig:raw_spoc_lc}
\end{figure*}

For the remainder of the system modeling in this work, we use the Simple Aperture Photometry (SAP) light curve from the SPOC pipeline \citep{twicken2010, morris2020} with the following pre-processing. First we removed contamination from nearby stars by scaling the light curves by the SPOC-provided \texttt{CROWDSAP} values. We also ignored data with nonzero SPOC quality flags. This included an anomalous event during Sector 13 (TJD 1665.2983 to 1665.3501) where the spacecraft fell out of fine pointing. The resulting light curve can be seen in Figure \ref{fig:raw_spoc_lc} with its Lomb-Scargle periodogram.

\begin{figure*}
    \centering
    \includegraphics[width=\textwidth]{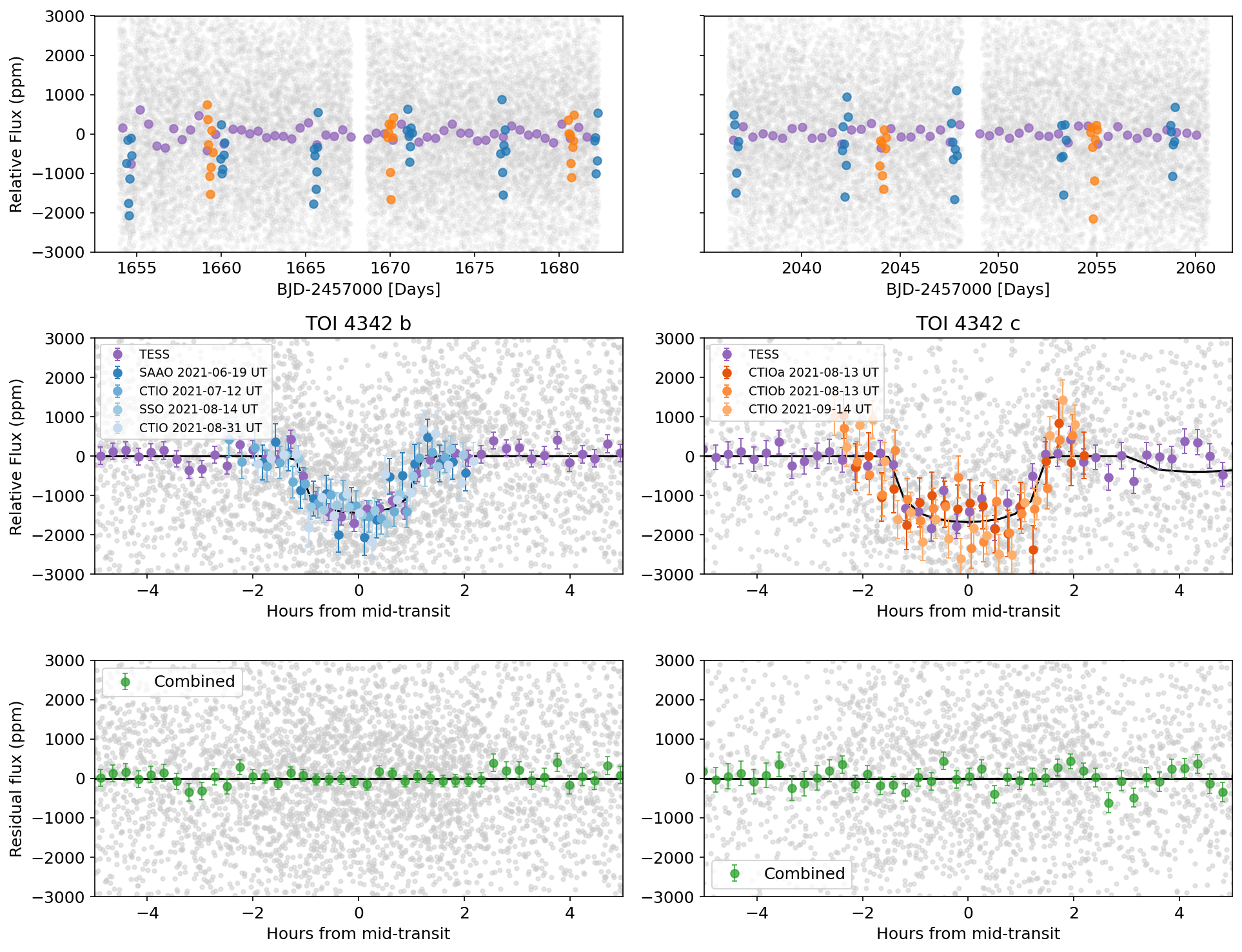}
    \caption{Top: Detrended \tess\ light curve in gray, binned values in purple, with transits highlighted for each \bPlanet\ (blue) and \cPlanet\ (orange). Middle: \tess\ and LCOGT light curves folded on the best fit period and epoch found in Section \ref{sec:bestfitmodel}. Different background-subtracted observations are shown in different colors for each transit with the best fit model in black. Bottom: Residual flux between the fully combined light curve and the model.}
    \label{fig:fit}
\end{figure*}

In the periodogram, we see a peak signifying stellar variability at a period of about 13 days. To remove this variability and other instrumental systematics, we conducted our own correction similar to our new QLP correction. First, we excluded data from QLP-predicted transit times, then we split the light curve into individual spacecraft orbits (with two orbits per sector) to be corrected separately. We again iteratively removed short-timescale systematics with quaternion time series statistics and long-timescale systematics with basis spline fits. This final fit was then divided out to produce our corrected light curve. Figure \ref{fig:fit} shows the detrended 2-minute cadence light curve with transits highlighted.

\subsection{Ground-based Photometry}
\label{sec:groundphot}

We obtained seeing-limited ground-based follow-up observations from the \tess\ Follow-up Observing Program Sub Group 1 \citep[TFOP SG1;][]{collins2019}. The ground-based observations have much higher spatial resolution than \tess\ and can help confirm the source location of a \tess\ transit signal and can provide additional transit observations to refine ephemerides for predicting future transits. We used the {\tt TESS Transit Finder}, which is a customized version of the {\tt Tapir} software package \citep{jensen2013}, to schedule our transit observations.

Between UT 2021-05-30 and 2021-09-14, 4 transits of \bPlanet\ and 5 transits of \cPlanet\ were observed in the Sloan $i'$ band using the Las Cumbres Observatory Global Telescope \citep[LCOGT;][]{brown2013} 1.0\,m network. The observations were taken at the Siding Spring Observatory (SSO), South Africa Astronomical Observatory (SAAO) and Cerro Tololo Inter-American Observatory (CTIO) nodes of the LCOGT network and are summarized in Table \ref{tab:sg1}. We use 7 of the 9 transits as 2 were cut short for bad weather conditions. The 1\,m telescopes are equipped with $4096\times4096$ SINISTRO cameras having an image scale of $0\farcs389$ per pixel, resulting in a $26\arcmin\times26\arcmin$ field of view. The images were calibrated by the standard LCOGT {\tt BANZAI} pipeline \citep{mccully2018}. Differential photometric data were extracted with {\tt AstroImageJ} \citep{collins2017} using circular photometric apertures with radii $6\farcs0$. Thus, the \target\ aperture excludes flux from the nearest known Gaia DR3 and TICv8 star (TIC 2025922721) $16\arcsec$ east of \target. As shown in Section \ref{sec:lcmodeling}, the transit signals are detected on-target relative to known Gaia DR3 stars. We also checked the light curves of Gaia DR3 sources within $2\farcm5$ of \target\ and found no evidence of nearby eclipsing binary systems that could be causing the \tess\ detection. \update{All SG1 data can be found online at \citet{https://doi.org/10.26134/exofop3}.}

\begin{deluxetable*}{lllllll}
\tablewidth{0pt}
\tablecaption{
  SG1 Follow-Up Observations
  \label{tab:sg1}
}
\tablehead{
  \colhead{Target} & \colhead{Instrument} & \colhead{Date (UT)} & \colhead{Filter} & \colhead{Aperture} &
  \colhead{Observing Notes}
}
\startdata
\cPlanet & LCO-CTIO 1.0m & 2021-09-14 & $i'$ & 5.1\arcsec &  \\
\bPlanet & LCO-CTIO 1.0m & 2021-08-31 & $i'$ & 6.2\arcsec &  \\
\bPlanet & LCO-SSO 1.0m & 2021-08-14 & $i'$ & 5.1\arcsec &  \\
\cPlanet & LCO-CTIO 1.0m a & 2021-08-13 & $i'$ & 5.9\arcsec & Simultaneous observation \tablenotemark{a} \\
\cPlanet & LCO-CTIO 1.0m b & 2021-08-13 & $i'$ & 5.9\arcsec & Simultaneous observation \tablenotemark{a} \\
\bPlanet & LCO-CTIO 1.0m & 2021-07-12 & $i'$ & 6.2\arcsec &  \\
\cPlanet & LCO-SAAO 1.0m & 2021-07-11 & $i'$ & 7.0\arcsec & Noisy, cut short by weather \\
\bPlanet & LCO-SAAO 1.0m & 2021-06-19 & $i'$ & 8.6\arcsec &  \\
\cPlanet & LCO-CTIO 1.0m & 2021-05-30 & $i'$ & 7.8" & Partial, cut short by weather \\
\enddata
\tablenotetext{a}{Transit observed simultaneously by two distinct telescopes at CTIO.}
\end{deluxetable*}

\subsection{Reconnaissance Spectroscopy} 
\label{sec:spec}

We obtained 9 spectra of \target\ over two seasons in slicer mode with the fiber-fed high resolution echelle spectrograph CHIRON \citep{chiron}. CHIRON is mounted on the 1.5\,m SMARTS telescope, located at the CTIO, Chile, and has a spectral resolving power of 80,000. The spectra were taken using 1 hour long exposures and were extracted by the standard CHIRON pipeline \citep{paredes2021}.

We derived the radial velocities using a cross correlation against a median combined template spectrum. The template spectrum is composed of a median combination of all CHIRON spectra, each shifted to rest after an approximate velocity measurement via a cross correlation against a synthetic template. The measured velocity of each spectrum is that of the mean velocity from each spectral order, weighted by their cross correlation function heights. The velocity uncertainties were estimated from the scatter of the per-order velocities. We find a mean internal uncertainty of $\sim$ 15 \ms, with an RMS of $\sim$ 22 \ms\ between the 9 measurements. The full dataset can be found in Table \ref{tab:rvs}.

\begin{deluxetable}{lrrl}
\tablewidth{0pc}
\tablecaption{
  Radial Velocities of \target 
  \label{tab:rvs}
}
\tablehead{
  \colhead{BJD} & \colhead{RV} & \colhead{$\sigma_{RV}$} & \colhead{Instrument} \\
  \colhead{} & \colhead{(\kms)} & \colhead{(\kms)} & \colhead{}
}
\startdata
$2459359.91837$ & $-5.726$ & $0.013$ & CHIRON \\
$2459360.85475$ & $-5.760$ & $0.018$ & CHIRON \\
$2459379.77461$ & $-5.795$ & $0.020$ & CHIRON \\
$2459384.82649$ & $-5.770$ & $0.011$ & CHIRON \\
$2459409.75314$ & $-5.788$ & $0.012$ & CHIRON \\
$2459700.86421$ & $-5.734$ & $0.018$ & CHIRON \\
$2459725.87319$ & $-5.747$ & $0.016$ & CHIRON \\
$2459740.83584$ & $-5.737$ & $0.016$ & CHIRON \\
$2459742.79459$ & $-5.761$ & $0.012$ & CHIRON \\
\enddata
\end{deluxetable}

\subsection{High-Resolution Speckle Imaging} 
\label{sec:speckle}

``Third-light” flux contamination from a close stellar companion can lead to an underestimated planetary radius if not accounted for in the transit model \citep{ciardi2015} and even cause non-detections of small planets residing within the same exoplanetary system \citep{lester2021}. The discovery of close, bound companion stars, which exist in nearly one-half of FGK type stars \citep{matson2018} and less so for M class stars, provides crucial information toward our understanding of exoplanetary formation, dynamics and evolution \citep{howell2021}. Thus, to search for close-in bound companions unresolved by \tess, Gaia, or ground-based seeing-limited follow-up observations, we obtained high-resolution imaging speckle observations of \target.

\begin{figure}
    \centering
    \includegraphics[width=.45\textwidth]{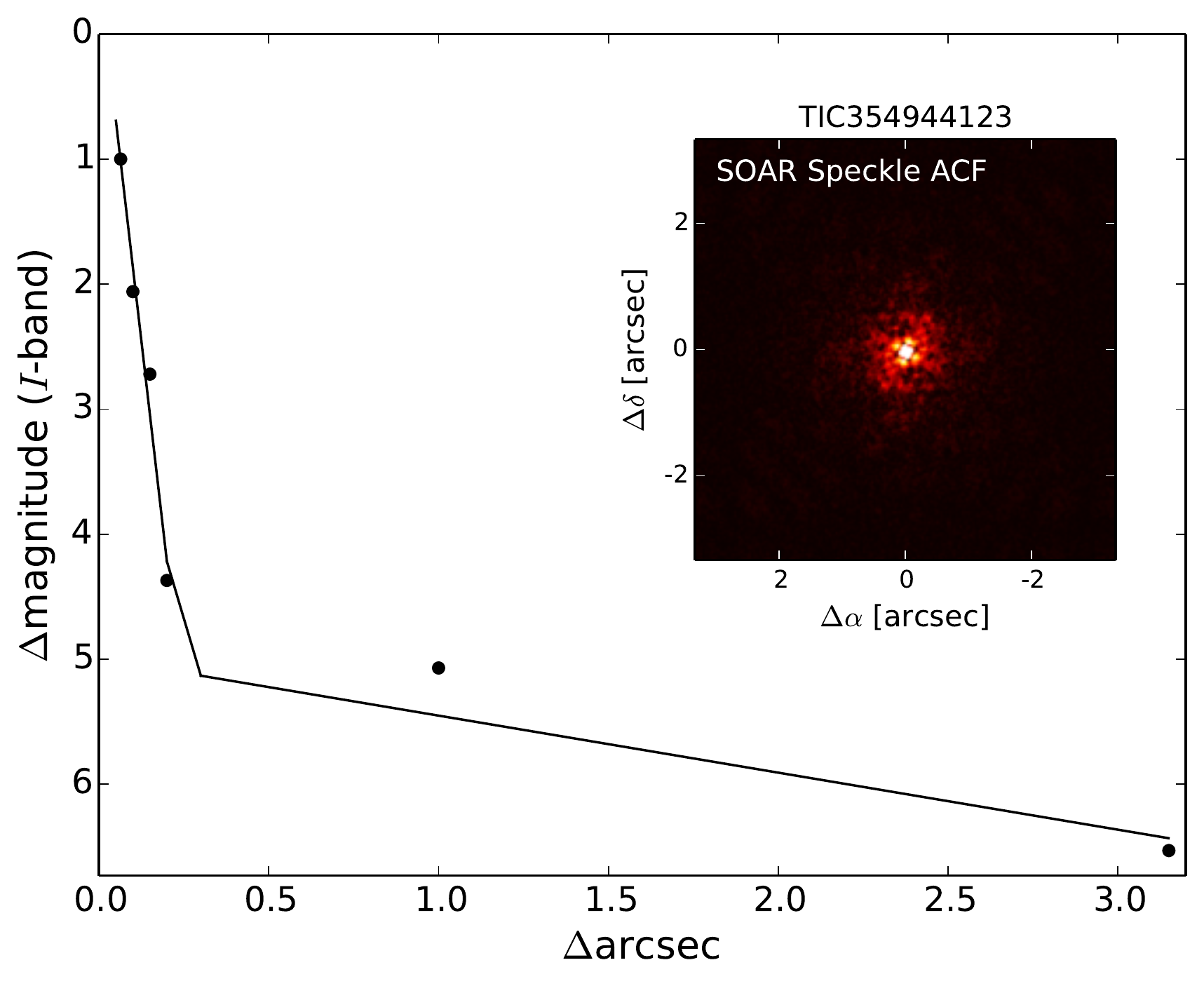}
    \caption{\update{SOAR speckle auto-correlation function (inset plot) and its 5 sigma detection sensitivity curve (main plot) in the Cousins-I band.}}
    \label{fig:soar}
\end{figure}

\target\ was observed at the 4.1-m Southern Astrophysical Research (SOAR) telescope \citep{soar} on 1 October 2021 UT, in Cousins I-band, a similar visible bandpass as \tess. This observation was sensitive to a 5-magnitude-fainter star at an angular distance of 1\arcsec\ from the target. More details of the observations within this survey are available in \citet{ziegler2020}. The $5\sigma$ detection sensitivity and speckle auto-correlation functions from the observations are shown in Figure \ref{fig:soar}. No nearby stars were detected within 3\arcsec\ of \target\ in the SOAR observations.

\begin{figure}
    \centering
    \includegraphics[width=.45\textwidth]{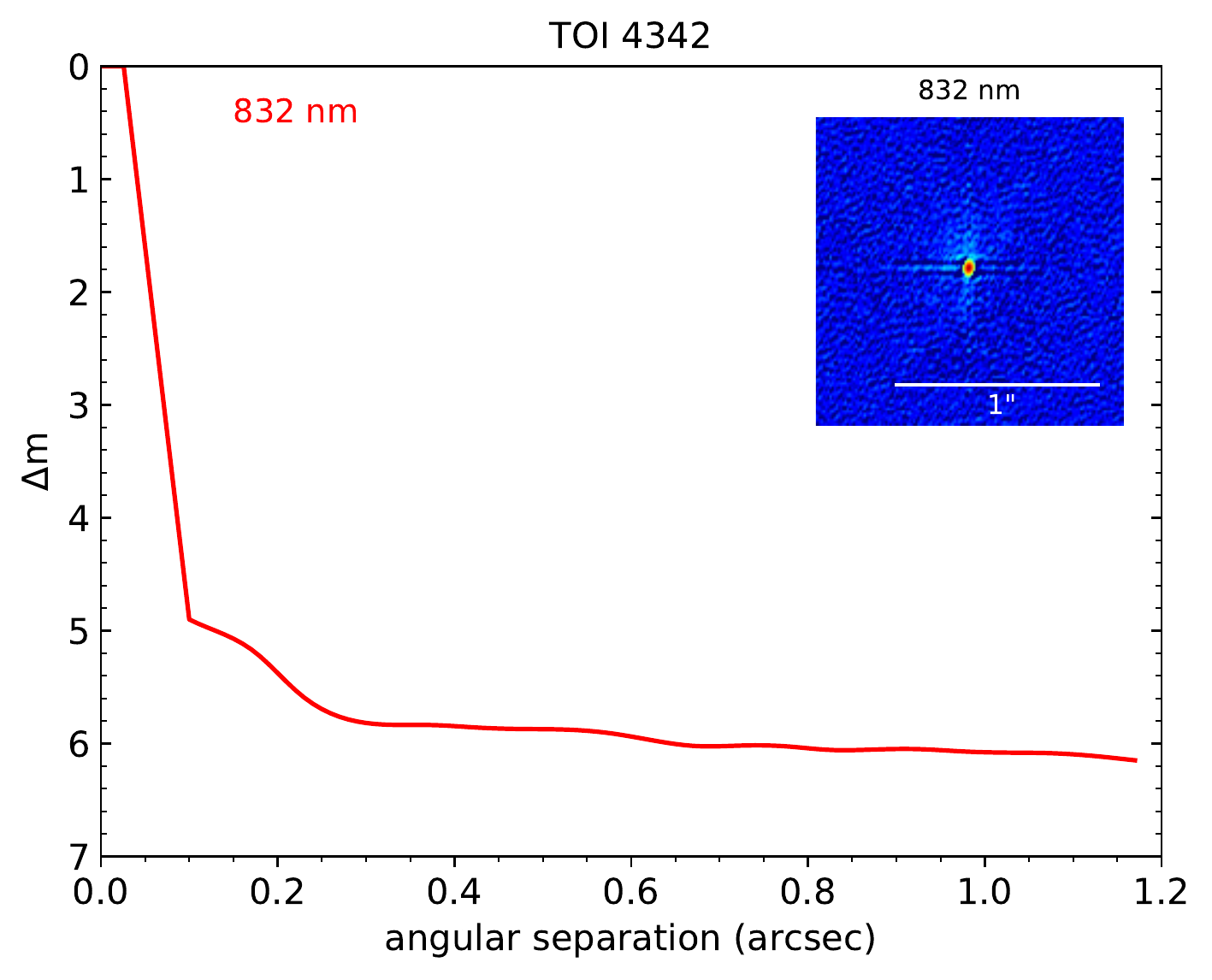}
    \caption{Gemini South Zorro Speckle Imaging 5 sigma contrast curve (full plot) and reconstructed images (inset plot) in the 832\,nm band. The diffraction limit of the instrument is 20\,mas.}
    \label{fig:gemini}
\end{figure}

\target\ was also observed on July 23 2021 UT using the Zorro speckle instrument on the Gemini South 8-m telescope\footnote{\url{https://www.gemini.edu/sciops/instruments/alopeke-zorro/}} \citep{alopeke-zorro}. Zorro provides simultaneous speckle imaging in two bands (562 nm and 832 nm) with output data products including a reconstructed image with robust contrast limits on companion detections \citep[e.g.,][]{howell2016}. During this observing run, the blue channel was inoperable, thus only 832 nm observations were obtained. Thirteen sets of 1000 X 0.06 sec exposures were collected and subjected to Fourier analysis in the standard reduction pipeline \citep[see][]{howell2011}. Figure \ref{fig:gemini} shows our final contrast curve and the 832 nm reconstructed speckle image. We find that \target\ is a single star with no companion brighter than 5-6 magnitudes below that of the target star from very close in (0.1\arcsec) out to 1.2\arcsec. At the distance of \target\ (\starApproximateDistance\ pc) these angular distances correspond to spatial distances of 6.2 to 74 au.

\section{Analysis}
\label{sec:analysis}
\subsection{Stellar parameters}
\label{sec:stellar_parameters}

We re-derive stellar parameters according to the following empirical relations. First we take the observed $K_s$
magnitude from 2MASS and the Gaia Data Release 3 (DR3) \citep{gaiacollaboration2022} parallax to find the absolute $K_s$ magnitude
of \target. Then, using the relation from \cite{benedict2016}, we get a stellar mass of \starMass\ \msun.
With this mass, we use the mass-radius relation from \cite{boyajian2012} to find a stellar radius of 
\starRadius\ \rsun. We check this with the absolute $K_s$ magnitude-radius relation from \cite{mann2015} 
where we find a consistent value of \starRadiusMann\ \rsun. From \citet{mann2015}, we also calculate a bolometric luminosity of
\starLuminosity\ \lsun\ via the observed V and J magnitudes and the resulting bolometric correction. Using the 
Stefan-Boltzmann law, we find an effective temperature of \starTeff\ K, giving \target\ a spectral type around M0V. These values are consistent with the stellar parameters listed in TIC 8.2 of \ticMass\ \msun, \ticRadius\ \rsun, \ticLuminosity\ \lsun, and \ticTeff\ K, \update{which were derived using the older Gaia DR2}.

\subsection{Light curve modeling} 
\label{sec:lcmodeling}

\subsubsection{Best Fit Model}
\label{sec:bestfitmodel}

We simultaneously fit the detrended \tess\ light curve and all 7 of the 9 SG1 light curves unaffected by weather (see Table \ref{tab:sg1}) using \texttt{exoplanet} \citep{exoplanet}. 

For \tess\ light curves, we approximate the per point measured uncertainty as 1.4826 times the median absolute deviation (MAD) of the flux within each \tess\ orbit. For each SG1 light curve, we use the reported flux uncertainties, and on top of that, fit for a jitter term (added in quadrature) to capture additional errors in the observation's error budget. We also simultaneously fit a second order polynomial to the SG1 light curves to account for nightly trends. To model the stellar limb darkening we use quadratic limb darkening models with uninformative priors following \cite{kipping2013} for each observation band.

For both orbits, we assume an eccentricity of 0. \footnote{\update{Based on priors from \citet{Kipping_2013} and \citet{Eylen_2019}, we expect the eccentricities to be low. We also performed a separate fit with an uninformed prior on the eccentricity and argument of periapsis, and did not see significant changes in the results.}} Periods and epochs for each planet were given uniform priors centered on the values found via BLS search to $\pm 10\%$ of the period. The ratios of planet to stellar radii had uniform priors from 0 to 1, and impact parameters had uniform priors from $-(1 + R_p / R_*)$ to $+(1 + R_p / R_*)$ (though when reported, we take the absolute value). Finally, the stellar mass and radius were given normal priors using the parameters derived in Section \ref{sec:stellar_parameters}.

\texttt{exoplanet} uses \texttt{PyMC3} \citep{pymc3} to perform No U-turn Sampling \citep{hoffman2011} from the posterior distribution. We sampled five independent chains with 5000 tuning steps and 5000 draws. All parameters converged with Gelman-Rubin statistic $\leq 1.01$ \citep{gelman1992}.  Table \ref{tab:fit} shows the median sampled and derived values with 1 sigma confidence intervals, and Figure \ref{fig:fit} shows the median posterior model with residuals.

\subsubsection{Transit Shape Model}
\label{sec:transitshapemodel}

Separately, using \texttt{exoplanet}'s \texttt{SimpleTransitOrbit} we also performed a \tess-only fit to best constrain each candidate's transit shape -- the ratio between the duration of the flat part of the transit ($t_F$) to total transit duration ($t_T$). 

Again, we assume a Gaussian noise model on top of a constant baseline. Priors on period and epoch were set to uniform distributions centered around the BLS periods and epochs with bounds of $\pm 10\%$ of the period. Durations were given uniform priors from 0 to 2 times the BLS durations, $\rpl / \rstar$ were given uniform priors from 0 to 1, and impact parameters were given uniform priors from $-(1 + R_p / R_*)$ to $+(1 + R_p / R_*)$. Since we want to find the best characterization of the transit shape for each planet independently from the host star, we use two independent star models with loose stellar radius priors from 0 to 2 \rsun\ and quadratic limb darkening with \citet{kipping2013}'s priors.

Following the same fitting configuration as Section \ref{sec:bestfitmodel}, we can calculate the transit shape following Equation 15 from \citep{seager2003}.
\begin{equation}
\left(\frac{t_F}{t_T}\right)^2 = \frac{(1 - R_p / R_*)^2 - b^2}{(1 + R_p / R_*)^2 - b^2} \ 
\end{equation}
where $R_p$ is planet radius, $R_*$ is stellar radius, and $b$ is impact parameter. For \bPlanet\ and \cPlanet\ respectively, we find median and one sigma confidence intervals of  \bTransitShape\ and \cTransitShape.

\subsection{Radial Velocity Modeling}
\label{sec:rv}
Using the CHIRON data (Table \ref{tab:rvs}), we can place mass upper limits on both planets by fitting for the amplitudes of simple sinusoids assuming circular orbits. This puts an upper bound on the radial velocity variations, meaning we can constrain any transiting companion masses to be planetary rather than stellar. 

We define our model with three unknowns: the baseline radial velocity, $K_b$, and $K_c$, where the $K$'s are semi-amplitudes of sinusoids set to the median periods and epochs from Section \ref{tab:fit}. After fitting, we find semi-amplitudes of \bSemiAmplitude\ \ms\ and \cSemiAmplitude\ \ms\ for each signal respectively. These posteriors give us 3-$\sigma$ mass upper limits of \bUpperMSinI\ \mjup\ and \cUpperMSinI\ \mjup\ both much smaller than stellar masses.

\subsection{Photocenter motion} 
\label{sec:diffimage}

Photocenter motion analysis can help determine if the location of a transit signal matches the location of the target star on the sky. 

An effective method of determining both of these locations is with the difference imaging technique \citep{bryson2013}, designed initially for the Kepler Mission and inherited by the SPOC pipeline, whereby the difference of averaged in- and out-of-transit pixel images is found. Assuming stellar variability and/or instrumental systematics are negligible on transit timescales, the difference image should appear star-like at the location of the transit signal source. Meanwhile, the out-of-transit image should represent a direct image of the field surrounding the target star. If the field is relatively uncrowded and the target star is indeed the source of the transit signal, the difference and direct images should appear similar.

To produce difference and direct images, we used \texttt{TESS-plots}\footnote{\url{https://github.com/mkunimoto/TESS-plots}}, a publicly available Python package for robust pixel-level analysis of \tess\ FFIs. For planets in multi-planet systems such as \target, \texttt{TESS-plots} masks out all cadences corresponding to other planet transits. This ensures that the primary source of variability in the difference image is due to the planet of interest. \texttt{TESS-plots} also puts more care in choosing which in- and out-of-transit frames are used compared to the original difference images in QLP. It ignores the first and last 5\% of the transit duration to better avoid ingress / egress. It also places a buffer between in- and out-of-transit frames to handle underestimated transit durations. Finally, ``bad" transits (transits with lots of missing or poor quality datapoints) are discarded to prevent contamination of the overall difference image. Altogether, with these improvements, \texttt{TESS-plots} marks an important update over the Primary Mission QLP difference images.

\begin{figure*}
    \centering
    \includegraphics[width=\textwidth]{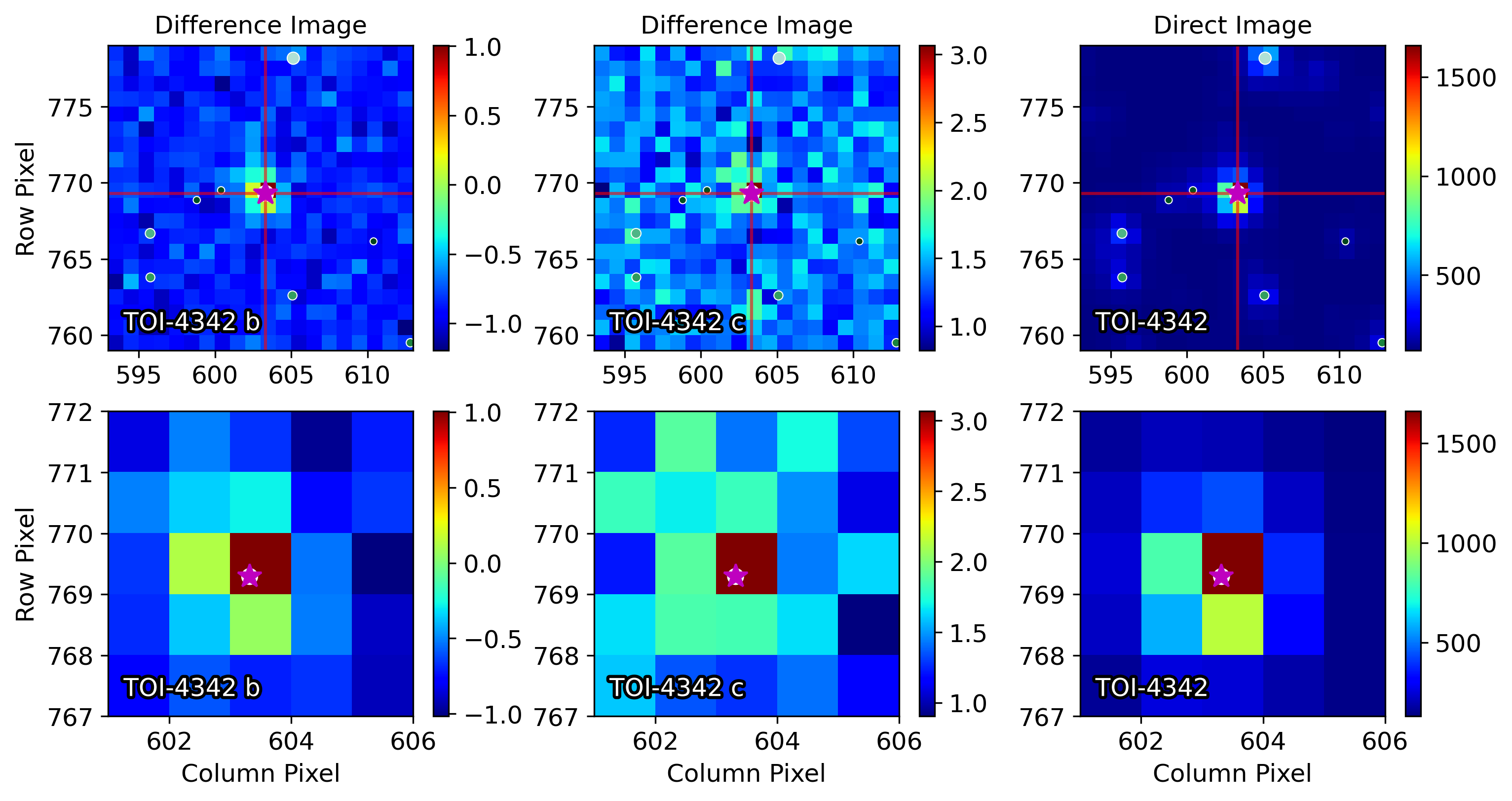}
    \caption{Difference images for \bPlanet\ and \cPlanet, made using 20×20 pixel cutouts of the Sector 27 \tess\ FFIs (top). A close-up of the central 5×5 pixels is also shown (bottom). The third column shows the average of out-of-transit images, which should represent a direct image of the field near the target star. The target TIC-354944123 is indicated with a pink star, while nearby stars down to $\Delta T = 4$ mag are plotted as white circles with sizes scaled by brightness. The difference images for both planet candidates indicate the transit sources are co-located with the target star. The color bars are all in units of electrons per second.}
    \label{fig:diffimage}
\end{figure*}

Figure \ref{fig:diffimage} shows difference images for \bPlanet\ and \cPlanet\ next to the direct image for \target, using Sector 27 FFIs. The difference images confidently rule out the transit signals as coming from other nearby sources listed in the TIC, consistent with the centroid analysis derived by the SPOC pipeline. For \bPlanet, the SPOC-derived photo centroid is only 2.5 $\sigma$ away from the location of TIC 2025922721 (T=19.805 mag). However, this star is too faint to produce the transit depths observed on \target.

\subsection{Possible False Positive Scenarios}
\label{sec:fps}

In this section we rule out possible false positive scenarios where the signals are not coming from a multi-planet system.

\subsubsection{\target\ is an eclipsing star system}
\label{sec:ebs}

One possible source of false positives could be signals from eclipsing stellar companions rather than planetary companions. We can rule out this scenario by considering our radial velocity model in Section \ref{sec:rv}. We found 3-$\sigma$ upper limits on semi-amplitude magnitudes of \bUpperSemiAmplitude\ \ms\ and \cUpperSemiAmplitude\ \ms\ which translate to mass limits of \bUpperMSinI\ \mjup\ and \cUpperMSinI\ \mjup\ for \bPlanet\ and \cPlanet\ respectively. In other words, if our transit signals are caused by gravitationally bound companions that block out light from \target, those companions must have sub-Jupiter masses and cannot be from stars.

\subsubsection{Contamination from a Nearby Eclipsing Binary}
\label{sec:nebs}

Another major source of false positives is the signal of an eclipsing binary (EB) in the field near our target of interest. Because photometers measure all light within a specific aperture, eclipses from a nearby EB (NEB) can contaminate the target aperture and cause transit-like events in the light curve. These false positives account for as much as 40\% of transit-like signals at the lowest Galactic latitudes in the Kepler field \citep{Morton2011, bryson2013}.

We can start ruling out NEBs by restricting the signal source to be near \target. Our photocenter motion analysis in Section \ref{sec:diffimage} rules out signals from known TICv8 stars, constraining the signal to be within $\sim 21 \arcsec$ (one \tess\ pixel) of \target. In Section \ref{sec:groundphot}, SG1 observations rule out signals from the nearest Gaia DR3 stars. In DR3 \citep{gaiacollaboration2022, fabricius2021}, Gaia has a resolution down to $\sim 0.7 \arcsec$, so the signal must be on-target or from an NEB within an arcsecond of \target.

Next, we can rule out potential NEBs by showing they must be brighter than certain magnitudes to cause either transit signal. First, we note the observed transit depth $\delta_{\text{obs}}$ is

%\begin{align*}
%    \dobs &= \frac{F_{\text{EB, out}} - F_{\text{EB, in}}}{F_{\text{EB, out}} + F_*} \\
%    &= \frac{F_{\text{EB, out}} - F_{\text{EB, in}}}{F_{\text{EB, out}}} \cdot \frac{F_{\text{EB, out}}}{F_{\text{EB, out}} + F_*} \\
%    &= \deb \cdot \frac{f}{1+f}
%\end{align*}

\begin{equation} \label{eq:dobs}
 \dobs = \deb \cdot \frac{f}{1+f}
\end{equation}

where $\deb$ the ``true" depth (the NEB's primary eclipse depth if \target\ were not present), and $f$ is the flux ratio between the NEB and \target.
%$f \equiv F_{\text{EB, out}} / F_*$ the flux ratio between the NEB and the star \target.  
%$F_*$ is the flux from \target, and $F_{\text{EB, in}}$ and $F_{\text{EB, out}}$ are the NEB's flux in and out of an eclipse, making  and .

Then, we can place an upper bound on $\deb$ by assuming $b=0$ as in Eq. 21 from \citet{seager2003}:

\begin{equation} 
\deb \leq \frac{(1 - \frac{t_F}{t_T})^2}{(1 + \frac{t_F}{t_T})^2}
\end{equation}

Together with Equation \ref{eq:dobs}, this places a lower bound on $f$ purely as a function of $\dobs$ and transit shape ($\frac{t_F}{t_T}$):

\begin{equation}
   \frac{\dobs / \deb}{1 - \dobs / \deb} \leq f 
\end{equation}

% $$\dobs \leq \frac{(1 - \frac{t_F}{t_T})^2}{(1 + \frac{t_F}{t_T})^2} \cdot \frac{f}{1+f}$$

which we can rearrange to an upper bound on NEB magnitude ($m_{\text{EB}} - m_* = -2.5 \log_{10}f$ where $m_*$ is the \starApproximateTMag, the $T_{\rm mag}$ of \target).

% Using the result of the transit shape model in Section \ref{sec:transitshapemodel}, we derive 2-$\sigma$ lower bounds on transit shape of \bTwoSigmaLowerTransitShape\ and \cTwoSigmaLowerTransitShape\ for each signal. These correspond to NEB magnitude upper bounds of \bTwoSigmaBlendMag\ and \cTwoSigmaBlendMag\ ($\Delta T = 2.83$ and $3.90$ mag). At 3-$\sigma$, these become $\Delta T = 4.35$ and $5.67$. In other words, if an NEB caused the \bPlanet\ signal, there's a 95\% chance it is within brighter than \bTwoSigmaBlendMag\ and a 99.7\% chance it's brighter than \bThreeSigmaBlendMag.

Using the result of the transit shape model in Section \ref{sec:transitshapemodel}, we can derive 3-$\sigma$ lower bounds on transit shape for each signal. These correspond to NEB magnitude upper bounds of \bThreeSigmaBlendMag\ and \cThreeSigmaBlendMag\ ($\Delta T = 4.35$ and $5.67$ mag). In other words, for an NEB to cause the \bPlanet\ signal, it is likely within 4.35 mag of \target.

In Figure \ref{fig:soar}, we see no stars within 5-6 mag of \target\ at separations of 1\arcsec\ -- 3.0\arcsec. This supports the ruling out of Gaia sources down to around 1\arcsec. In Figure \ref{fig:gemini}, no neighbors within 5-6 mag of \target\ were detected from 0.1\arcsec\ -- 1.2\arcsec. So, altogether we can rule out NEBs greater than 0.1\arcsec away from \target\ as sources of the transit signals.

Lastly, taking advantage of the relatively high proper-motion of \target, we can use archival images to rule out NEBs within 0.1\arcsec of \target's current location. Figure \ref{fig:field} shows the field surrounding \target\ in 1976 along with all known nearby TIC stars for magnitude reference. We see that in 1976, the 2021 location of \target\ is clear of any stars with $T_{\text{mag}} \lesssim 17$ ($\Delta T \simeq 6$ mag). This rules out NEBs within 0.1\arcsec of \target.

Altogether, using our photocenter motion analysis, high-resolution speckle imaging, and archival field images, we are able to rule out contamination from an NEB as the source of either transit signal.

\subsubsection{\target\ is a hierarchical triple}

The final scenario we consider is an EB gravitationally bound to \target\ (i.e. a hierarchical triple system). This EB would cause the same aperture contamination described in Section \ref{sec:nebs} (with the same magnitude limits), but could have evaded detection in Figure \ref{fig:field} because it would stay close to \target.

In this scenario, if we assume the EB's orbit is within 0.1 arcsec of \target, its semimajor axis would be at most $\sim 6.2$ AU (at a distance of \starApproximateDistance\ pc via Gaia DR3 parallax). With Kepler's third law and the stellar mass from Section \ref{sec:stellar_parameters}, this gives the NEB an orbital period of $\sim$ 19.2 years around \target.

From Section \ref{sec:nebs}, we saw that NEBs should be brighter than $\Delta T \simeq 5$ mag to cause the transit signals. This translates to a minimum luminosity of $7.5 \times 10^{-4} \lsun$ and, using $L \propto M^{3.5}$, a minimum mass of $0.13 \msun$.

For a gravitationally bound NEB with a mass of at least $0.13 \msun$, we would expect an edge-on radial velocity semiamplitude of $\sim$ 1.8 \kms. Gaia DR3 observed \target\ over a 34 month baseline. In this timeframe, if there were a 0.13 \msun\ companion at 6.2 AU, we would expect an RV shift of $\sim 1$ \kms. The reported \update{mean RV error} though is only $\sim 0.36$ \kms. 
Similarly, from Section \ref{sec:rv}, our RV data spans more than a year. Based on this observation timeline, we expect an RV scatter from a companion NEB \update{at a distance of 6.2 AU with $0.13 \msun$} would be larger than our observed 22 \ms\ at least 95\% of the time. 

Together, since the actual RV error is lower than expected from a companion EB for both our RV data and Gaia data, we conclude that these scenarios of EBs graviationally bound to \target\ are unlikely causes of the transit signals.

\subsubsection{TRICERATOPS \& Summary}
Using two sectors of \tess\ data along with additional photometric and spectroscopic observations, we were able to rule out most astrophysical false positives as sources of our transit signals. We considered the possibilities that our signals are caused directly by an EB, by contamination from a background EB, and by certain configurations of a hierarchical companion EB. For each scenario, we showed that it is highly unlikely for that scenario to cause our transits. In addition, we use \texttt{triceratops} \citep{triceratops} to independently check the false positive probabilities (FPPs) and nearby false positive probabilities (NFPPs) for each signal. After 20 runs for each signal, we calculate mean and standard deviation FPPs of: \bTriceratopsFPP\ and \cTriceratopsFPP\ and NFPPs of: \bTriceratopsNFPP\ and \cTriceratopsNFPP. 
% This means both satisfy the validated planet criteria of FPP $< 0.015$ and NFPP $< 10^{-3}$ as set in \citet{triceratops}. 
Finally, \citet{lissauer2012} and \citet{toi} found that multi-candidate systems have \textit{lower} false positive rates than single-candidate systems, so we receive a ``multiplicity boost", further decreasing the false positive probabilities and increasing the likelihood of having real planets. Altogether, we conclude our signals are statistically valid exoplanet transits.

\section{Results and Discussion}
\label{sec:discussion}
% \begin{itemize}
%     \item on the smaller end of semimajor axis + distance to earth -> good for atmosphere study
%     \item unclear what atmosphere based on temp/radius (could've been stripped away) mention that the outer planet is slightly larger than the inner planets with lower irradiation. Would be interesting to find out if the different irradiation is the driven factor for the light radius difference. 
%     \item short period -> stronger signals + better mass
%     \item conclusion
% \end{itemize}

In this work, we statistically validated a pair of sub-Neptunes around M0V dwarf \target. In this section, we discuss this system in the context of other planetary systems.

\subsection{Mass and Atmosphere Follow-Up Characterization}
The best fit parameters from Section \ref{sec:bestfitmodel} show the planets have radii of \bRadius\ and \cRadius\ \rearth, and periods of \bApproximatePeriod\ and \cApproximatePeriod\ days. Given their radii, these planets are most likely sub-Neptunes with a significant fraction of H/He in their atmospheres \citep{rogers2015}. Using sub-Neptune mass-radius relationship from \citet{wolfgang2016}, we expect the planets to have masses of \bMass\ and \cMass\ \mearth. 

This corresponds to expected radial velocity semi-amplitudes of \bApproximatePredictedK\ and \cApproximatePredictedK\ \ms, meaning it is feasible to measure the precise masses using a high precision radial velocity instrument mounted on a large telescope for \target\ ($V=\starApproximateVMag$ mag). With these masses, we will be able to compare bulk densities of the planets. Given the radii similarity (within 10\% of each other), this will show the influence of different levels of irradiation (incident fluxes of $\sim 27$ and $\sim 11$ $S_\earth$) on the planet atmospheres. 

Using the expected masses, we can also calculate transmission spectroscopy metrics \citep[TSMs,][]{kempton2018}, measures of how promising planets are for atmospheric characterization studies. For each planet, we find values of \bApproximateTSM\ and \cApproximateTSM, both of which are above the updated follow-up threshold of $\sim 25$ recommended by \citet{toi} for the 100 best atmospherically characterizable sub-Neptunes (updated from \citet{kempton2018}).

% Using the best fit parameters from Section \ref{sec:bestfitmodel}, we can calculate the transmission spectroscopy metric (TSM) as defined in \citet{kempton2018} to understand how promising these planets are for atmospheric characterization studies. For \bPlanet\ and \cPlanet, we find TSM values of \bTSM\ and \cTSM\ respectively. Notably, these are both larger than $\sim$ 30, making both planets among the best small \tess\ sub-Neptunes for atmospheric follow-up \citep{toi}. 

\begin{figure*}
    \centering
    \includegraphics[width=.75\textwidth]{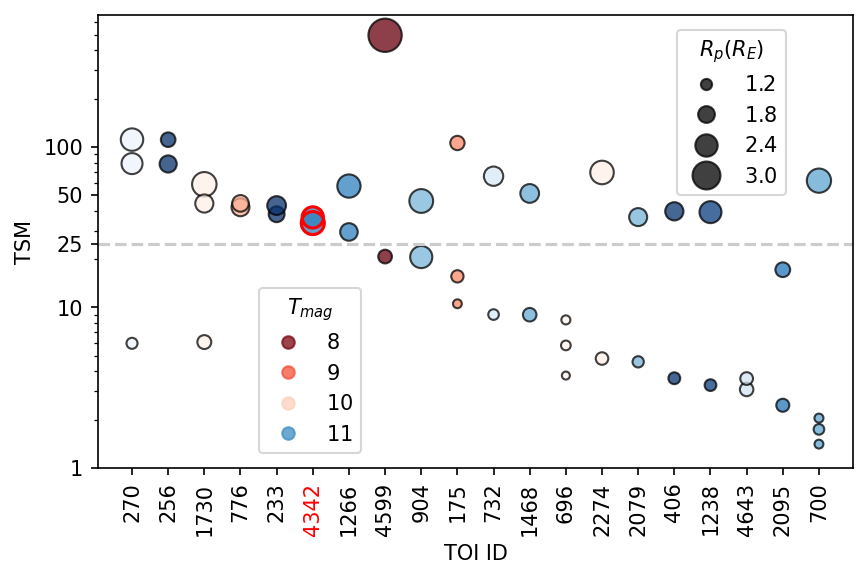}
    \caption{TSM, as defined in \citet{kempton2018}, for TOIs in multi-planet systems around bright M-dwarfs ($\teff < 4000$ K and $T_\text{mag} < 11.5$) with $\rpl < 4 \rearth$ as of July 7, 2022. TICs are sorted decreasingly by 2nd-largest TSM value in the system. Color corresponds to host $T_\text{mag}$ and size corresponds to planet radius. The gray dashed line is the atmospheric follow-up threshold recommended by \citet{toi} for sub-Neptunes. \target\ is among the top 10 systems with multiple planets that are well-suited for atmospheric characterization.}
    \label{fig:tsm}
\end{figure*}

More notably, \target\ has \textit{multiple} (more than one) high-TSM planets, making it one of the best M-dwarf systems for atmospheric comparison studies. Figure \ref{fig:tsm} shows the TSM values for multi-planet M-dwarf systems sorted by second highest TSM value of planets in each system. Characterizing and comparing the atmospheres of both planets will allow us to perform comparative exoplanetology, and \target\ is one of the few systems where multiple planets have characterizable atmospheres. Additionally, given the two planets are near a mean-motion resonance (MMR) of 2:1, it's likely they migrated together to their current orbits and have similar primordial compositions. This would mean any differences detected in their atmospheric properties can probably be attributed to the differing levels of stellar irradiation between the planets. This will help us gain insights on planetary atmosphere evolution and responses to different intensities of stellar irradiation.

\subsection{Small Planet Radius Gap}
\begin{figure*}[h]
    \centering
    \includegraphics[clip, trim=1cm 1cm 1cm 1cm, width=\textwidth]{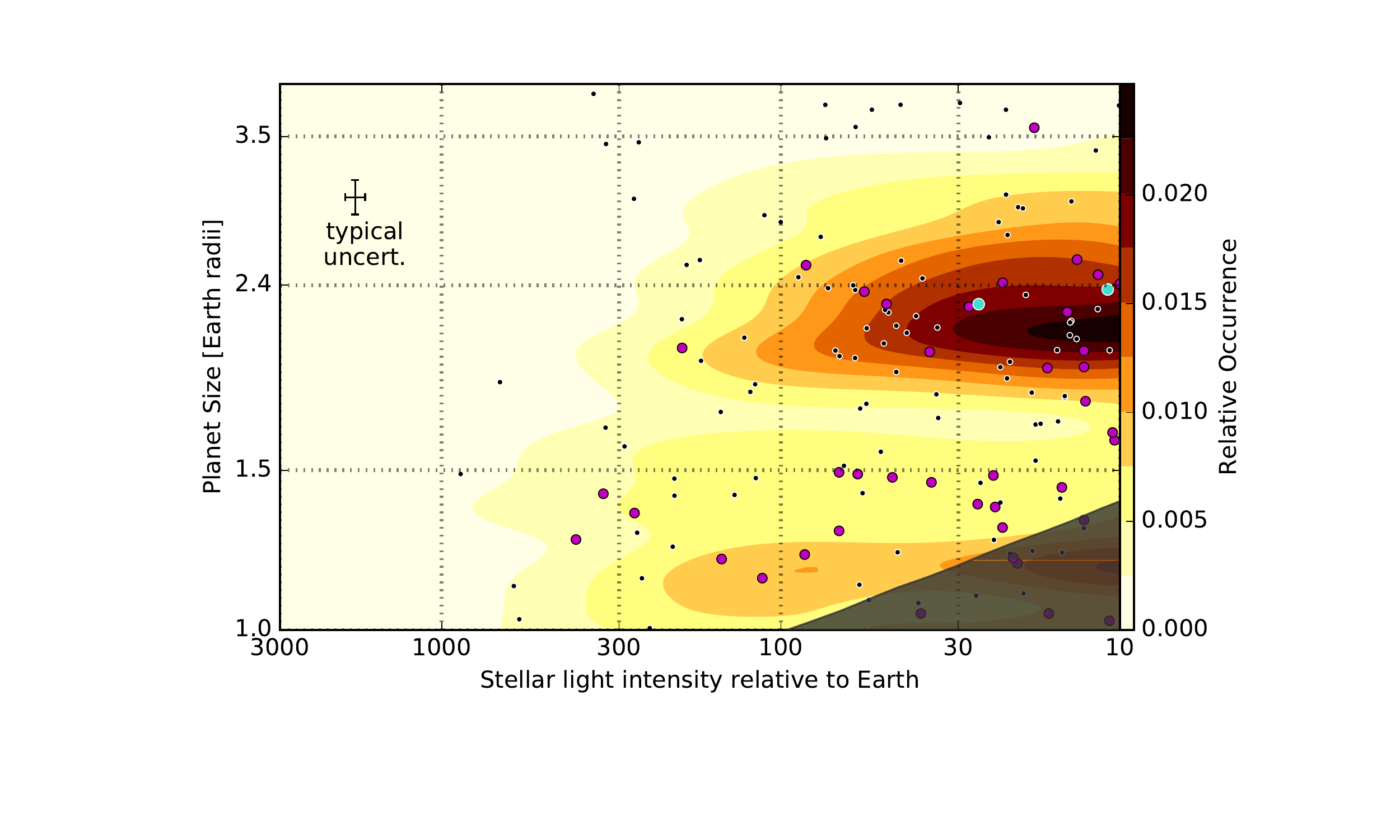}
    \caption{Insolation and orbital period vs planet size. \bPlanet\ and \cPlanet\ are shown in cyan over occurence contours for host stars with $M_* < 0.97 \msun$ from \citet{fulton2018}. M-dwarf TOIs as of July 7, 2022 are plotted as well, with multi-planet system planets highlighted in magenta. \bPlanet\ and \cPlanet\ both lie in the upper mode of the radius valley.
    % In the period vs planet size plot, estimates of the radius valley are shown from \citet{cloutier2020} (blue) and \citet{toi-270:vaneylen2021} (green). \bPlanet\ and \cPlanet\ both lie in the upper mode of the radius valley.
    }
    \label{fig:gap}
\end{figure*}

\citet{fulton2017} identified a radius gap for small planets roughly between 1.5 and 2.0 \rearth\ separating rocky super-Earths and gaseous sub-Neptunes for sun-like stars. \citet{cloutier2020} showed this gap persisted around low-mass stars. One predominant cause of this gap may be photoevaporation: the stripping away of a planet's atmosphere as it undergoes heavy irradiation from its star \citep{owen2013}. Highly irradiated planets would be left as bare rocky cores, while less irradiated planets would keep their atmospheres with larger mass and radii, leading to a bimodal radius distribution. With radii of \bApproximateRadius\ and \cApproximateRadius\ \rearth, both \bPlanet\ and \cPlanet\ appear to fall on the upper mode of the valley. Figure \ref{fig:gap} shows each \bPlanet\ and \cPlanet\ in cyan as a function of irradiation level and planetary radius over relative occurrence contours from \citet{fulton2018}. We see both planets have relatively low irradiation levels of $\sim 27$ and $\sim 11$ $S_\earth$ respectively, meaning they are likely good fits for the low mass atmosphere-retaining sub-Neptune planet description on the upper side of the gap.

\subsection{Transit Timing Variation}
\label{sec:ttv}
With periods of \bApproximatePeriod\ days and \cApproximatePeriod\ days, \bPlanet\ and \cPlanet\ also fall within 5\% of the first order MMR of 2:1. Given the close distance to the 2:1 resonance, we expect the system would show transit timing variation (TTV) signals. Based on formulas in \citet{Lithwick2012}, the super period of the TTV is about 157 days, the amplitude of the TTV signal is expected to be on the order of a few minutes using the estimated mass from empirical relations. Calculations using TTVFast \citep{ttvfast} assuming the expected planet masses and eccentricities smaller than 0.1 for both planets show similar results to \citet{Lithwick2012}. For \target, the photometric observations from ground-based 1-meter telescopes were able to achieve transit time measurements at a similar or slightly better precision than TESS. With these observations, we can search for evidence of TTVs over a baseline of $>800$ days via \texttt{exoplanet}'s \texttt{TTVOrbit}. First, we subtract the best fit background trends found in Section \ref{sec:bestfitmodel} from each light curve. We use the global best fit ephemerides to set normal priors on transit times with standard deviations of 7 minutes. The rest of the parameters (limb darkening, stellar properties, radii ratios, impact parameter) are initialized following the best fit model. Sampling parameters similarly followed the best global fit model. Based on the TTV fit, we do not observe significant deviation from linear ephemerides by more than 5 mins for a majority of the observations, indicating that the eccentricities of both planets are likely to be close to zero. \tess\ will reobserve \target\ in Sectors 66 \& 67 (June -- July 2023), adding new transits that will help further constrain the TTV amplitudes. Additional photometry observations that can achieve transit center timing with precision better than 1 min can also provide stronger constraints on the TTV amplitudes. If measured, these amplitudes could help derive constrain planet masses.
%Figures \ref{fig:resonance} and \ref{fig:resonance2} show the system compared to other multi-planet systems.

\begin{figure*}
    \centering
    \includegraphics[width=\textwidth]{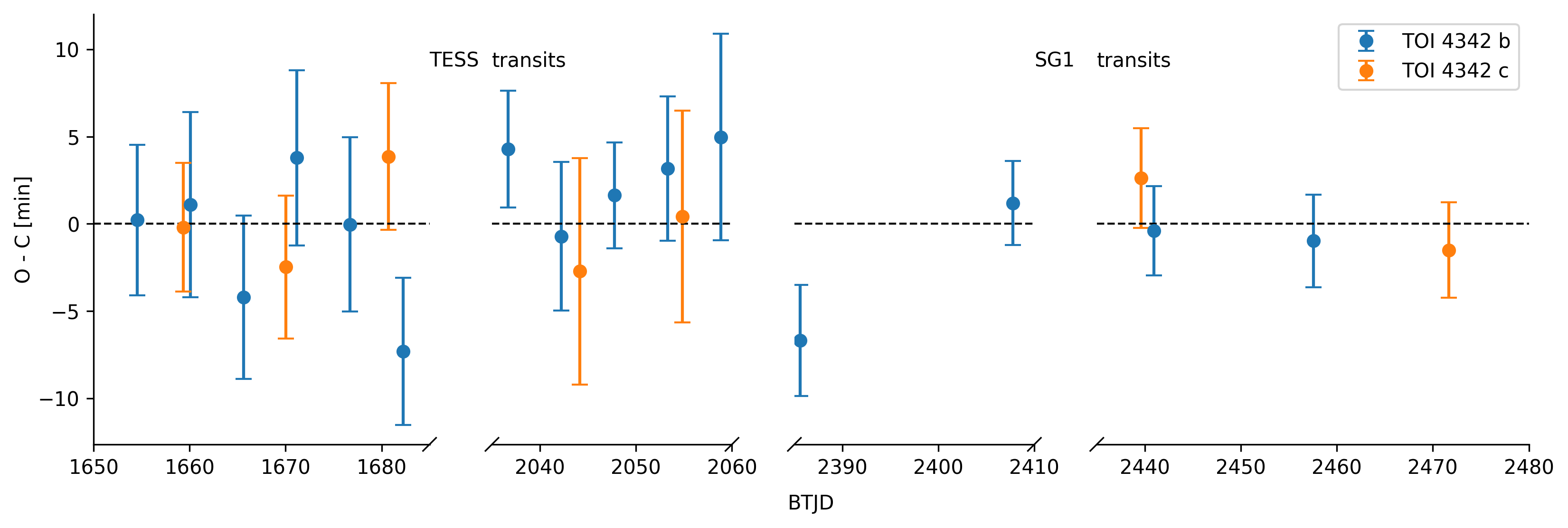}
    \caption{Observed minus calculated transit times for all observed transits across \tess\ and SG1 data. Observed transit times were modeled with \texttt{exoplanet}'s \texttt{TTVOrbit} (Section \ref{sec:ttv}) and the expected transit times were linearly propagated from the best fit ephemerides (Section \ref{sec:bestfitmodel}). Based on the current data, we see no evidence for significant deviations from expected transit times. With expected TTV amplitudes of only $\sim 5$ minutes, future observations could help could pinpoint TTV amplitudes.} \ref{sec:discussion})
    \label{fig:ttv}
\end{figure*}

% Commented out -- we'll discuss in text instead.
% \begin{figure*}
%     \centering
%     \includegraphics[width=.5\textwidth]{figures/resonance.png}
%     \caption{Caption}
%     \label{fig:resonance}
% \end{figure*}

% \begin{figure*}
%     \centering
%     \includegraphics[width=.5\textwidth]{figures/resonance2.png}
%     \caption{Caption}
%     \label{fig:resonance2}
% \end{figure*}

\subsection{QLP and Extended Mission FFIs}

This discovery showcases our recent improvements to QLP on Extended Mission FFIs. 
By adding a multi-planet search, short-timescale systematic correction, and improved difference images to QLP, we were able to detect this new M-dwarf system using the 1st Extended Mission FFIs. We saw signal-to-noise improvements when comparing transit searches between the original detrending method and the improved method. We also saw improvements when comparing searches between the Primary Mission and Extended Mission FFIs. On top of this, the multi-planet search, together with the longer light curve baseline, let us discover \textit{both} planets in the system, where the original QLP would only have found one. Finally, we were able to use the improved difference images in localizing the source of the transit events to our particular target.

\target\ is just one example of systems we will be able to find with all these improvements. In the future, we expect these upgrades to yield even more multi-planet M-dwarf systems as they continue to be used with every new sector of the standard QLP planet detection procedures at MIT. This will help build up our populations of small planets that are suitable for follow-up.

% \begin{acknowledgements}
% There's a problem where line numbers are forced when
% the acknowledgement environment is used with amsmath
% https://github.com/AASJournals/AASTeX60/issues/130
\section*{Acknowledgements}

% % TESS and MAST
This paper includes data collected by the \tess{} mission, which are publicly available from the Mikulski Archive for Space Telescopes (MAST) \citep{https://doi.org/10.17909/t9-yk4w-zc73, https://doi.org/10.17909/t9-9j8c-7d30, https://doi.org/10.17909/t9-r086-e880}).
Funding for the \tess{} mission is provided by NASA's Science Mission directorate.
We acknowledge the use of public TESS data from pipelines at the TESS Science Office and at the TESS Science Processing Operations Center.
% % SPOC
Resources supporting this work were provided by the NASA High-End Computing (HEC) Program through the NASA Advanced Supercomputing (NAS) Division at Ames Research Center for the production of the SPOC data products.
% % EXOFOP
This research has made use of the Exoplanet Follow-up Observation Program website \citep{https://doi.org/10.26134/exofop3}, which is operated by the California Institute of Technology, under contract with the National Aeronautics and Space Administration under the Exoplanet Exploration Program.
% % Exoplanet archive
This research has made use of the NASA Exoplanet Archive, which is operated by the California Institute of Technology, under contract with the National Aeronautics and Space Administration under the Exoplanet Exploration Program.

% % gemini south from steve howell 10/18
Some of the observations in the paper made use of the High-Resolution Imaging instrument Zorro obtained under Gemini LLP Proposal Number: GN/S-2021A-LP-105. Zorro was funded by the NASA Exoplanet Exploration Program and built at the NASA Ames Research Center by Steve B. Howell, Nic Scott, Elliott P. Horch, and Emmett Quigley. Zorro was mounted on the Gemini South telescope of the international Gemini Observatory, a program of NSF’s OIR Lab, which is managed by the Association of Universities for Research in Astronomy (AURA) under a cooperative agreement with the National Science Foundation. on behalf of the Gemini partnership: the National Science Foundation (United States), National Research Council (Canada), Agencia Nacional de Investigación y Desarrollo (Chile), Ministerio de Ciencia, Tecnología e Innovación (Argentina), Ministério da Ciência, Tecnologia, Inovações e Comunicações (Brazil), and Korea Astronomy and Space Science Institute (Republic of Korea).

% % LCO
This work makes use of observations from the LCOGT network. Part of the LCOGT telescope time was granted by NOIRLab through the Mid-Scale Innovations Program (MSIP). MSIP is funded by NSF.

% %Authors
% CXH acknowledges support from MIT's Kavli Institute as a Torres postdoctoral fellow and from Australian Research Council as a DECRA Fellow. 
% AV's work was performed under contract with the California Institute of Technology / Jet Propulsion Laboratory funded by NASA through the Sagan Fellowship Program executed by the NASA Exoplanet Science Institute.

%{\it Facilities:} 
\facility{TESS,
Gaia,
CTIO:1.5m (CHIRON),
LCO:1.0m (Sinistro),
SOAR:4.1m (HRCam),
Gemini South:8m (Zorro)}

\software{
    AstroImageJ \citep{astroimagej},
    Astropy \citep{astropy2013, astropy2018},
    \textsf{exoplanet} \citep{exoplanet, agol2020, arviz, kipping2013, starry, pymc3, theano},
    H5py,
    Matplotlib \citep{matplotlib},
    MIT Quick Look Pipeline \citep{qlp},
    Numpy \citep{numpy},
    \tess{} SPOC Pipeline \citep{spoc, li2018, twicken2018},
    Pandas \citep{pandas},
    Scipy \citep{scipy},
    Vartools \citep{vartools}. 
}

% \end{acknowledgements}

\bibliographystyle{apj}
\bibliography{refs}

\begin{deluxetable*}{lcr}

\tablewidth{0pc}
\tabletypesize{\scriptsize}
\tablecaption{
    Stellar and Planet Parameters for \target
    \label{tab:fit}
}
\tablehead{
    \multicolumn{1}{c}{~~~~~~~~Parameter~~~~~~~~} &
    \multicolumn{1}{c}{Value}                     &
    \multicolumn{1}{c}{Source}    
}
\startdata
\noalign{\vskip -3pt}
\sidehead{Catalog Information}
~~~~R.A. (h:m:s)                      & \starRA     & Gaia DR3\\
~~~~Dec. (d:m:s)                      & \starDec    & Gaia DR3\\
~~~~Epoch							  & \starRefEpoch   & Gaia DR3 \\
~~~~Parallax (mas)                    & \starParallax  & Gaia DR3\\
~~~~$\mu_{ra}$ (mas yr$^{-1}$)        & \starPMRA   & Gaia DR3 \\
~~~~$\mu_{dec}$ (mas yr$^{-1}$)       & \starPMDec & Gaia DR3\\
~~~~Gaia DR3 ID                       & \starGaiaID   &  \\
~~~~TIC ID                            & \starTICID & \\
~~~~TOI ID                            & \starTOIID  & \\
\sidehead{Photometric properties}
~~~~$TESS$ (mag)\dotfill            & \starTMag  & TIC v8.2         \\
~~~~$Gaia$ (mag)\dotfill            & \starGaiaMag & Gaia DR3               \\
~~~~Gaia RP (mag)\dotfill          & \starGaiaRPMag & Gaia DR3                 \\
~~~~Gaia BP (mag)\dotfill          & \starGaiaBPMag & Gaia DR3                 \\
~~~~$V_J$ (mag)\dotfill             & \starVMag & UCAC4 \tablenotemark{a}\\
~~~~$B_J$ (mag)\dotfill             & \starBMag & APASS DR9 \tablenotemark{b}\\
~~~~$J$ (mag)\dotfill               & \starJMag & 2MASS           \\
~~~~$H$ (mag)\dotfill               & \starHMag & 2MASS           \\
~~~~$K_s$ (mag)\dotfill             & \starKMag & 2MASS           \\
\sidehead{Derived properties}
~~~~$\mstar$ ($\msun$)\dotfill      &  \starMass & Parallax +\citet{benedict2016}\tablenotemark{c}\\
~~~~$\rstar$ ($\rsun$)\dotfill      & \starRadius & Parallax +\citet{mann2015} \tablenotemark{d}        \\
~~~~$\loggstar$ (cgs)\dotfill       & \starLogg &  empirical relation + LC \tablenotemark{e}        \\
~~~~$\lstar$ ($\lsun$)\dotfill      & \starLuminosity  & \citet{mann2015}     \\
~~~~$\teffstar$ (K)\dotfill        &  \starTeff  & \tablenotemark{f}\\
~~~~$M_V$ (mag)\dotfill &  $14.39\pm$0.02  & Parallax         \\
~~~~$M_K$ (mag)\dotfill &  $8.272\pm$0.015  & Parallax         \\
~~~~Distance (pc)\dotfill           & \starDistance  & Parallax\\
~~~~\rhostar (\gcmc)\dotfill &  \starRho & empirical relation + LC \tablenotemark{e} \\
% \enddata
% \end{deluxetable*}
%}
%\startdata
%\noalign{\vskip -3pt}
\sidehead{Limb-darkening coefficients}
~~~$u_1,TESS$               \dotfill    & \starTESSuOne     &  \\
~~~$u_2,TESS$               \dotfill    &  \starTESSuTwo    & \\
~~~$u_1,i'$               \dotfill    & \starIpuOne     & \\
~~~$u_2,i'$               \dotfill    &  \starIpuTwo     & \\  
[1.5ex]
\Lc\ parameters & \textbf{\bPlanet}\ & \textbf{\cPlanet}\ \\
~~~$P$ (days)  \dotfill    &  \bPeriod            & \cPeriod \\
~~~$T_c$ (${\rm BJD} - 2457000$)   \dotfill    &   \bEpoch   & \cEpoch         \\
~~~$T_{14}$ (hr) \dotfill    &   \bDuration &  \cDuration       \\
~~~$T_{12} = T_{34}$ (min)   \dotfill    &  \bIngressDuration  & \cIngressDuration      \\
~~~$\arstar$              \dotfill & \bAOR & \cAOR \\
~~~$\rpl/\rstar$          \dotfill & \bROR & \cROR \\
~~~$b \equiv a \cos i/\rstar$ \dotfill    & \bImpactParameter   & \cImpactParameter \\
~~~$i$ (deg) \dotfill &  \bInclination  & \cInclination \\
\sidehead{Planetary parameters}
~~~$\rpl$ ($\rearth$) \dotfill &   \bRadius   & \cRadius \\
~~~$a$ (AU) \dotfill & \bSemimajorAxis  &  \cSemimajorAxis \\
~~~$T_{\rm eq}$ (K) \dotfill &   \bTeq & \cTeq \\
~~~$\langle F \rangle$ ($S_\earth$) \dotfill    & \bIrr &  \cIrr 
\enddata

\tablenotetext{a}{\citet{zacharias2013}}
\vspace{-.25cm}
\tablenotetext{b}{\citet{henden2016}}
\vspace{-.25cm}
\tablenotetext{c}{We adopt error based on the scatter
in the empirical relations from \citet{benedict2016}}
\vspace{-.25cm}
\tablenotetext{d}{We adopt error based on the scatter
in the empirical relations from \citet{mann2015}}
\vspace{-.25cm}
\tablenotetext{e}{We fitted the transit light curves with a prior constraint on the stellar density.}
\vspace{-.25cm}
\tablenotetext{f}{$T_{\text{eff}}$ was determined from the bolometric luminosity and the stellar radius.}

\end{deluxetable*}
\end{document}

%% file: macros.tex
\usepackage{xcolor}

\newcommand{\target}{TOI 4342}
\newcommand{\bPlanet}{TOI 4342 b}
\newcommand{\cPlanet}{TOI 4342 c}

\newcommand{\rsun}{\ensuremath{R_\sun}}
\newcommand{\msun}{\ensuremath{M_\sun}}
\newcommand{\lsun}{\ensuremath{L_\sun}}

\newcommand{\rearth}{\ensuremath{R_\earth}}
\newcommand{\mearth}{\ensuremath{M_\earth}}

\newcommand{\mjup}{\ensuremath{M_{\rm Jup}}}

\newcommand{\teff}{\ensuremath{T_{\rm eff}}}

\newcommand{\rpl}{\ensuremath{R_{p}}}

\newcommand{\kms}{\ensuremath{\rm km\,s^{-1}}}
\newcommand{\ms}{\ensuremath{\rm m\,s^{-1}}}

\newcommand{\rstar}{\ensuremath{R_\star}}
\newcommand{\mstar}{\ensuremath{M_\star}}
\newcommand{\lstar}{\ensuremath{L_\star}}

\newcommand{\teffstar}{\ensuremath{T_{\rm eff\star}}}
\newcommand{\rhostar}{\ensuremath{\rho_\star}}
\newcommand{\loggstar}{\ensuremath{\log{g_{\star}}}}

\newcommand{\arstar}{\ensuremath{a/\rstar}}

\newcommand{\tess}{{\it TESS}}

\newcommand{\Lc}{Light curve\ }

\newcommand{\gcmc}{\ensuremath{\rm g\,cm^{-3}}}

\newcommand{\C}{\ensuremath{^{\circ}C\;}}

\newcommand{\dobs}{\delta_{\text{obs}}}
\newcommand{\deb}{\delta_{\text{EB}}}

\input{fit_params.tex}

%% file: fit_params.tex
\newcommand{\starTOIID}{\ensuremath{4342}}
\newcommand{\starTICID}{\ensuremath{354944123}}
\newcommand{\starGaiaID}{\ensuremath{6355915466181029376}}
\newcommand{\starRA}{21:37:33.48}
\newcommand{\starDec}{-77:58:44.9743}
\newcommand{\starRefEpoch}{\ensuremath{2016.0}}
\newcommand{\starPMRA}{\ensuremath{120.333\pm0.019}}
\newcommand{\starPMDec}{\ensuremath{-91.503\pm0.018}}

\newcommand{\starParallax}{\ensuremath{16.249\pm0.018}}
\newcommand{\starTMag}{\ensuremath{11.0318\pm0.0074}}
\newcommand{\starGaiaMag}{\ensuremath{11.97403\pm0.00032}}
\newcommand{\starGaiaBPMag}{\ensuremath{12.8963\pm0.0014}}
\newcommand{\starGaiaRPMag}{\ensuremath{11.02247\pm0.00081}}
\newcommand{\starJMag}{\ensuremath{9.832\pm0.024}}
\newcommand{\starHMag}{\ensuremath{9.179\pm0.024}}
\newcommand{\starKMag}{\ensuremath{9.018\pm0.021}}
\newcommand{\starVMag}{\ensuremath{12.669\pm0.057}}
\newcommand{\starBMag}{\ensuremath{14.055\pm0.011}}

\newcommand{\ticMass}{\ensuremath{0.587\pm0.020}}
\newcommand{\ticRadius}{\ensuremath{0.598\pm0.018}}
\newcommand{\ticTeff}{\ensuremath{3880\pm160}}
\newcommand{\ticLuminosity}{\ensuremath{0.073\pm0.016}}

\newcommand{\starDistance}{\ensuremath{61.543\pm0.070}}
\newcommand{\starMass}{\ensuremath{0.6296\pm0.0086}}
\newcommand{\starRadius}{\ensuremath{0.599\pm0.013}}
\newcommand{\starRadiusMann}{\ensuremath{0.598\pm0.018}}
\newcommand{\starLuminosity}{\ensuremath{0.0746\pm0.0053}}
\newcommand{\starTeff}{\ensuremath{3901\pm69}}

\newcommand{\starApproximateDistance}{\ensuremath{61.54}}
\newcommand{\starApproximateTMag}{\ensuremath{11.032}}
\newcommand{\starApproximateVMag}{\ensuremath{12.67}}
\newcommand{\starApproximateMass}{\ensuremath{0.63}}
\newcommand{\starApproximateRadius}{\ensuremath{0.60}}
\newcommand{\starApproximateTeff}{\ensuremath{3900}}

\newcommand{\bQLPPeriod}{\ensuremath{5.538}}

\newcommand{\bQLPSNR}{\ensuremath{13.126}}
\newcommand{\cQLPPeriod}{\ensuremath{10.688}}

\newcommand{\cQLPSNR}{\ensuremath{11.506}}

\newcommand{\bSemiAmplitude}{\ensuremath{19.4_{-5.0}^{+5.1}}}
\newcommand{\bUpperSemiAmplitude}{\ensuremath{50}}
\newcommand{\cSemiAmplitude}{\ensuremath{14.4_{-3.6}^{+3.6}}}
\newcommand{\cUpperSemiAmplitude}{\ensuremath{36}}
\newcommand{\bUpperMSinI}{\ensuremath{0.321}}
\newcommand{\cUpperMSinI}{\ensuremath{0.289}}

\newcommand{\starRho}{\ensuremath{2.985_{-0.097}^{+0.089}}}
\newcommand{\starLogg}{\ensuremath{4.6878_{-0.0096}^{+0.0086}}}
\newcommand{\starTESSuOne}{\ensuremath{0.27_{-0.11}^{+0.13}}}
\newcommand{\starTESSuTwo}{\ensuremath{0.37_{-0.18}^{+0.17}}}
\newcommand{\starIpuOne}{\ensuremath{0.44_{-0.14}^{+0.13}}}
\newcommand{\starIpuTwo}{\ensuremath{0.08_{-0.17}^{+0.18}}}

\newcommand{\bEpoch}{\ensuremath{1654.53559_{-0.00061}^{+0.00059}}}
\newcommand{\bPeriod}{\ensuremath{5.5382498_{-0.0000058}^{+0.0000057}}}
\newcommand{\bROR}{\ensuremath{0.03491_{-0.00049}^{+0.00047}}}
\newcommand{\bMass}{\ensuremath{7.83_{-0.88}^{+0.93}}}
\newcommand{\bImpactParameter}{\ensuremath{0.290_{-0.050}^{+0.042}}}
\newcommand{\bSemimajorAxis}{\ensuremath{0.05251_{-0.00011}^{+0.00011}}}
\newcommand{\bAOR}{\ensuremath{18.97_{-0.21}^{+0.19}}}
\newcommand{\bInclination}{\ensuremath{89.13_{-0.13}^{+0.15}}}
\newcommand{\bRadius}{\ensuremath{2.266_{-0.038}^{+0.038}}}

\newcommand{\bTeq}{\ensuremath{633.6_{-6.3}^{+6.2}}}

\newcommand{\bIrr}{\ensuremath{26.9_{-1.0}^{+1.1}}}
\newcommand{\bDuration}{\ensuremath{2.215_{-0.018}^{+0.020}}}
\newcommand{\bIngressDuration}{\ensuremath{4.88_{-0.13}^{+0.15}}}

\newcommand{\cEpoch}{\ensuremath{1659.34623_{-0.00094}^{+0.00096}}}
\newcommand{\cPeriod}{\ensuremath{10.688716_{-0.000015}^{+0.000015}}}
\newcommand{\cROR}{\ensuremath{0.03719_{-0.00057}^{+0.00054}}}
\newcommand{\cMass}{\ensuremath{8.53_{-0.94}^{+0.90}}}
\newcommand{\cImpactParameter}{\ensuremath{0.187_{-0.061}^{+0.060}}}
\newcommand{\cSemimajorAxis}{\ensuremath{0.08140_{-0.00017}^{+0.00017}}}
\newcommand{\cAOR}{\ensuremath{29.40_{-0.32}^{+0.29}}}
\newcommand{\cInclination}{\ensuremath{89.63_{-0.12}^{+0.12}}}
\newcommand{\cRadius}{\ensuremath{2.415_{-0.040}^{+0.043}}}

\newcommand{\cTeq}{\ensuremath{508.9_{-5.0}^{+5.0}}}

\newcommand{\cIrr}{\ensuremath{11.18_{-0.44}^{+0.45}}}
\newcommand{\cDuration}{\ensuremath{2.820_{-0.025}^{+0.024}}}
\newcommand{\cIngressDuration}{\ensuremath{6.31_{-0.14}^{+0.16}}}

\newcommand{\bApproximatePeriod}{\ensuremath{5.538}}
\newcommand{\bApproximateRadius}{\ensuremath{2.27}}

\newcommand{\bApproximateTSM}{\ensuremath{36}}

\newcommand{\bApproximatePredictedK}{\ensuremath{3.9}}

\newcommand{\cApproximatePeriod}{\ensuremath{10.689}}
\newcommand{\cApproximateRadius}{\ensuremath{2.41}}

\newcommand{\cApproximateTSM}{\ensuremath{32}}

\newcommand{\cApproximatePredictedK}{\ensuremath{3.4}}

\newcommand{\bTransitShape}{\ensuremath{0.912_{-0.026}^{+0.013}}}

\newcommand{\bThreeSigmaBlendMag}{\ensuremath{15.38}}

\newcommand{\cTransitShape}{\ensuremath{0.907_{-0.039}^{+0.016}}}

\newcommand{\cThreeSigmaBlendMag}{\ensuremath{16.70}}

\newcommand{\bTriceratopsFPP}{\ensuremath{0.00211\pm0.00018}}
\newcommand{\bTriceratopsNFPP}{\ensuremath{0.0000138\pm0.0000013}}
\newcommand{\cTriceratopsFPP}{\ensuremath{0.00320\pm0.00027}}
\newcommand{\cTriceratopsNFPP}{\ensuremath{0.000387\pm0.000026}}

%% file: authors.tex
\author[0000-0002-5308-8603]{Evan~Tey}
\affiliation{Department of Physics and Kavli Institute for Astrophysics and Space Science, Massachusetts Institute of Technology, 77 Massachusetts Ave, Cambridge, MA, 02139, USA}

\author[0000-0003-0918-7484]{Chelsea~X.~Huang}
\affiliation{University of Southern Queensland, West St, Darling Heights, Toowoomba, Queensland, 4350, Australia}

\author[0000-0001-9269-8060]{Michelle~Kunimoto}
\affiliation{Department of Physics and Kavli Institute for Astrophysics and Space Science, Massachusetts Institute of Technology, 77 Massachusetts Ave, Cambridge, MA, 02139, USA}

\author[0000-0001-7246-5438]{Andrew~Vanderburg}
\affiliation{Department of Physics and Kavli Institute for Astrophysics and Space Science, Massachusetts Institute of Technology, 77 Massachusetts Ave, Cambridge, MA, 02139, USA}

\author[0000-0002-1836-3120]{Avi~Shporer}
\affiliation{Department of Physics and Kavli Institute for Astrophysics and Space Science, Massachusetts Institute of Technology, 77 Massachusetts Ave, Cambridge, MA, 02139, USA}

\author[0000-0002-8964-8377]{Samuel~N. ~Quinn}
\affiliation{Center for Astrophysics | Harvard \& Smithsonian, 60 Garden St, Cambridge, MA, 02138, USA}

\author[0000-0002-4891-3517]{George~Zhou}
\affiliation{University of Southern Queensland, West St, Darling Heights, Toowoomba, Queensland, 4350, Australia}

\author[0000-0001-6588-9574]{Karen~A. ~Collins}
\affiliation{Center for Astrophysics | Harvard \& Smithsonian, 60 Garden St, Cambridge, MA, 02138, USA}

\author[0000-0003-2781-3207]{Kevin~I.~Collins}
\affiliation{George Mason University, 4400 University Drive, Fairfax, VA, 22030, USA}

\author[0000-0002-4625-7333]{Eric~L. N. ~Jensen}
\affiliation{Department of Physics and Astronomy, Swarthmore College, 500 College Ave, Swarthmore, PA, 19081, USA}

\author[0000-0001-8227-1020]{Richard~P.~Schwarz}
\affiliation{Center for Astrophysics | Harvard \& Smithsonian, 60 Garden St, Cambridge, MA, 02138, USA}

\author{Ramotholo~Sefako}
\affiliation{South African Astronomical Observatory, P.O. Box 9, Observatory, Cape Town, 7935, South Africa}

\author[0000-0002-4503-9705]{Tianjun~Gan}
\affiliation{Department of Astronomy and Tsinghua Centre for Astrophysics, Tsinghua University, Beijing, 100084, China}

\author[0000-0001-9800-6248]{Elise~Furlan}
\affiliation{Caltech/IPAC-NExScI, NASA Exoplanet Science Institute, 1200 East California Boulevard, Mail Code 100-22, Pasadena, CA, 91125, USA}

\author[0000-0003-2519-6161]{Crystal~L.~Gnilka}
\affiliation{NASA Ames Research Center, Moffett Field, CA, 94035, USA}

\author[0000-0002-2532-2853]{Steve~B.~Howell}
\affiliation{NASA Ames Research Center, Moffett Field, CA, 94035, USA}

\author[0000-0002-9903-9911]{Kathryn~V.~Lester}
\affiliation{NASA Ames Research Center, Moffett Field, CA, 94035, USA}

\author{Carl~Ziegler}
\affiliation{Department of Physics, Engineering and Astronomy, Stephen F. Austin State University, 1936 North St, Nacogdoches, TX, 75962, USA}

\author{C\'{e}sar~Brice\~{n}o}
\affiliation{Cerro-Tololo Inter-American Observatory, Casilla 603, La Serena, Chile}

\author{Nicholas~Law}
\affiliation{Department of Physics and Astronomy, The University of North Carolina at Chapel Hill, Chapel Hill, NC, 27599-3255, USA}

\author[0000-0003-3654-1602]{Andrew~W.~Mann}
\affiliation{Department of Physics and Astronomy, The University of North Carolina at Chapel Hill, Chapel Hill, NC, 27599-3255, USA}

\author[0000-0003-2058-6662]{George~R. ~Ricker}
\affiliation{Department of Physics and Kavli Institute for Astrophysics and Space Science, Massachusetts Institute of Technology, 77 Massachusetts Ave, Cambridge, MA, 02139, USA}

\author[0000-0001-6763-6562]{Roland~K.~Vanderspek}
\affiliation{Department of Physics and Kavli Institute for Astrophysics and Space Science, Massachusetts Institute of Technology, 77 Massachusetts Ave, Cambridge, MA, 02139, USA}

\author[0000-0001-9911-7388]{David~W. ~Latham}
\affiliation{Center for Astrophysics | Harvard \& Smithsonian, 60 Garden St, Cambridge, MA, 02138, USA}

\author[0000-0002-6892-6948]{S.~Seager}
\affiliation{Department of Physics and Kavli Institute for Astrophysics and Space Science, Massachusetts Institute of Technology, 77 Massachusetts Ave, Cambridge, MA, 02139, USA}
\affiliation{Department of Earth, Atmospheric and Planetary Sciences, Massachusetts Institute of Technology, 77 Massachusetts Ave, Cambridge, MA, 02139, USA}
\affiliation{Department of Aeronautics and Astronautics, Massachusetts Institute of Technology, 77 Massachusetts Avenue, Cambridge, MA 02139, USA}

\author[0000-0002-4715-9460]{Jon~M.~Jenkins}
\affiliation{NASA Ames Research Center, Moffett Field, CA, 94035, USA}

\author{Joshua~N. ~Winn}
\affiliation{Department of Astrophysical Sciences, Princeton University, 4 Ivy Lane, Princeton, NJ, 08544, USA}

\author[0000-0003-1963-9616]{Douglas~A.~Caldwell}
\affiliation{NASA Ames Research Center, Moffett Field, CA, 94035, USA}
\affiliation{SETI Institute, Moffett Field, Mountain View, CA, 94035, USA}

\author[0000-0002-9003-484X]{David~Charbonneau}
\affiliation{Center for Astrophysics | Harvard \& Smithsonian, 60 Garden St, Cambridge, MA, 02138, USA}

% \author[0000-0002-5741-3047]{David~R.~Ciardi }
% \affiliation{Caltech/IPAC-NExScI, NASA Exoplanet Science Institute, 1200 East California Boulevard, Mail Code 100-22, Pasadena, CA, 91125, USA}

\author[0000-0002-7754-9486]{Christopher~J.~Burke}
\affiliation{Department of Physics and Kavli Institute for Astrophysics and Space Science, Massachusetts Institute of Technology, 77 Massachusetts Ave, Cambridge, MA, 02139, USA}

% \author[0000-0003-1286-5231]{David~R.~Rodriguez}
% \affiliation{Space Telescope Science Institute, 3700 San Martin Drive, Baltimore, MD, 21218, USA}

\author[0000-0002-2482-0180]{Zahra~Essack}
\affiliation{Department of Physics and Kavli Institute for Astrophysics and Space Science, Massachusetts Institute of Technology, 77 Massachusetts Ave, Cambridge, MA, 02139, USA}
\affiliation{Department of Earth, Atmospheric and Planetary Sciences, Massachusetts Institute of Technology, 77 Massachusetts Ave, Cambridge, MA, 02139, USA}